\documentclass[fleqn,usenatbib]{mnras}

\usepackage{newtxtext,newtxmath}

\usepackage[T1]{fontenc}
\usepackage{ae,aecompl}

\usepackage{graphicx}	
\usepackage{amsmath}	
\usepackage{csvtools}
\usepackage{multicol}
\usepackage[normalem]{ulem}
\usepackage{float}
\restylefloat{table}

\usepackage{subcaption}
\usepackage{float}

\usepackage{soul}
\newcommand\hii{H$\;${\small\rmfamily{II}}\relax}

\newcommand\hco{HCO$^{+}$\relax}
\newcommand\hcoi{HCO$^{+}$(1--0)\relax}

\newcommand\coi{$^{13}$CO(1--0)\relax}
\newcommand\coii{$^{13}$CO(2--1)\relax}

\title[Star formation in the S254-S258 region]{The link between gas and stars in the S254-S258 star-forming region}

\author[Dmitry A. Ladeyshchikov et al.]{D. A. Ladeyshchikov$^1$\thanks{Dmitry.ladeyschikov@urfu.ru}, M. S. Kirsanova$^{2}$, A. M. Sobolev$^1$, M. Thomasson$^3$, \newauthor
V. Ossenkopf-Okada$^4$, M. Juvela$^5$, S. A. Khaibrakhmanov$^1$, E. A. Popova$^1$\\
$^1$ Ural Federal University, 51 Lenin Str., Ekaterinburg 620051, Russia \\
$^2$ Institute of Astronomy, Russian Academy of Sciences, 48 Pyatnitskaya Str., Moscow  119017, Russia \\
$^3$ Department of Space, Earth and Environment, Chalmers University of Technology, Onsala Space Observatory, SE-43992 Onsala, Sweden \\
$^4$ I. Physik. Institut, University of Cologne, D-50937 Cologne, Germany\\
$^5$ University of Helsinki, Department of Physics, PL 64 (Gustaf H\"allstr\"omin katu 2), Finland
}

\date{Accepted XXX. Received YYY; in original form ZZZ}

\pubyear{2021}

\begin{document}
\label{firstpage}
\pagerange{\pageref{firstpage}--\pageref{lastpage}}
\maketitle

\begin{abstract}
The paper aims to study relation between the distributions of the young stellar objects (YSOs) of different ages and the gas-dust constituents of the S254-S258 star-formation complex. This is necessary to study the time evolution of the YSO distribution with respect to the gas and dust compounds which are responsible for the birth of the young stars. For this purpose we use correlation analysis between different gas, dust and YSOs tracers. We compared the large-scale CO, HCO$^+$, near-IR extinction, and far-IR {\it Herschel}  maps with the density of YSOs of the different evolutionary Classes. The direct correlation analysis between these maps was used together with the wavelet-based spatial correlation analysis. This analysis reveals a much tighter correlation of the gas-dust tracers with the distribution of Class~I YSOs than with that of Class~II YSOs. We argue that Class~I YSOs which were initially born in the central bright cluster S255-IR (both N and S parts) during their evolution to Class~II stage ($\sim$2 Myr) had enough time to travel through the whole S254-S258 star-formation region. Given that the region contains several isolated YSO clusters, the evolutionary link between these clusters and the bright central S255-IR (N and S) cluster can be considered. Despite the complexity of the YSO cluster formation in the non-uniform medium, the clusters of Class~II YSOs in the S254-258 star-formation region can contain objects born in the different locations of the complex.
\end{abstract}




\section{Introduction}

S254-S258 is a region that shows signs of sequential star formation triggered by expanding \hii{} regions (see, e.g., \citet{Bieging07,Minier07,Chavarria08, Ojha2011}). It contains two bright optical \hii{} regions S255 and S257 with angular sizes about 3\arcmin, an extended \hii{} region S254 with size about 10\arcmin{} and two compact regions S256 and S258 with sizes about 1\arcmin.   The region contains 10 young star clusters that were identified in the work of  \citet[]{Chavarria08} (see Figure 12 therein): S255-2 \& S255N, S256, S258, G192.54-0.15, G192.75-0.00, G192.75-0.08, G192.63-0.00, G192.69-0.25, G192.65-0.08, G192.55-0.01. At present, only the brightest molecular condensations with embedded clusters S255N, S255-IR, and S255S were mapped with high-density gas tracing molecular lines \citep{Cyganowski07, Wang11, Zinchenko12,Zinchenko15,Zemlyanukha18}. The expansion of the \hii{} regions S255 and S257 \citep[observed recently by e.~g.][]{buslaeva2021}  and compressing the material  "sandwiched" between these \hii{} regions may explain a large amount of high-density gas and active star formation in the S255-IR region \citep{Minier07}. Other molecular condensations with embedded clusters have not been studied yet in the high-density gas tracers. Most of them reside in the primordial gas (i.e. in the gas which has not been compressed by expanding \hii{} regions) of the S254-S258 star formation region mapped by \citet{Bieging07,Chavarria08}. Two of the condensations (S256-south and S258) are located towards the S256 and S258 \hii{} regions, while others, including G172.75-0.00, G192.75-0.08, and G192.69-0.25, are located on the periphery of the region.

The presence of the high-density gas in the position of stellar clusters can be checked by observations of e.g. CS and HCO$^+$ molecular line emission \citep{Gaches15}. The CS(2--1) line in the brightest condensations S255N and S255-IR was firstly detected by \citet{Zinchenko94} and was observed several times in the recent years \citep{Zinchenko09, Wang11, Zinchenko15} together with the \hco{} line \citep{Zinchenko09, Zinchenko15, Shirley13}. 
The aim of the present study is to map the CS(2--1) and \hcoi{} emission of the surrounding gas, including mapping of isolated star clusters in order to study its spatial structure and kinematics. We also observed optically thin lines C$^{18}$O(1--0), C$^{34}$S(2--1), and H$^{13}$CO$^+$(1--0) toward selected directions to reveal kinematics of the inner parts of the clumps with high extinction. We additionally compare different gas and dust tracers to the distribution of the young stellar objects (YSOs) density to reveal the connection between young star clusters and the surrounding gas using the correlation analysis method. Class I and class II YSOs represent different evolutionary stages of the objects with protoplanetary disks. So, these studies provide constraints on the distribution of exoplanet-hosting stars. 

\section{Observations and processing}  \label{sec:obs_proc}

\subsection{High-density molecules observations} \label{hight_density_molecules}

\begin{figure*}
\centering
\subcaptionbox{CS (2--1) integrated intensity}{\includegraphics [width=0.80\linewidth] {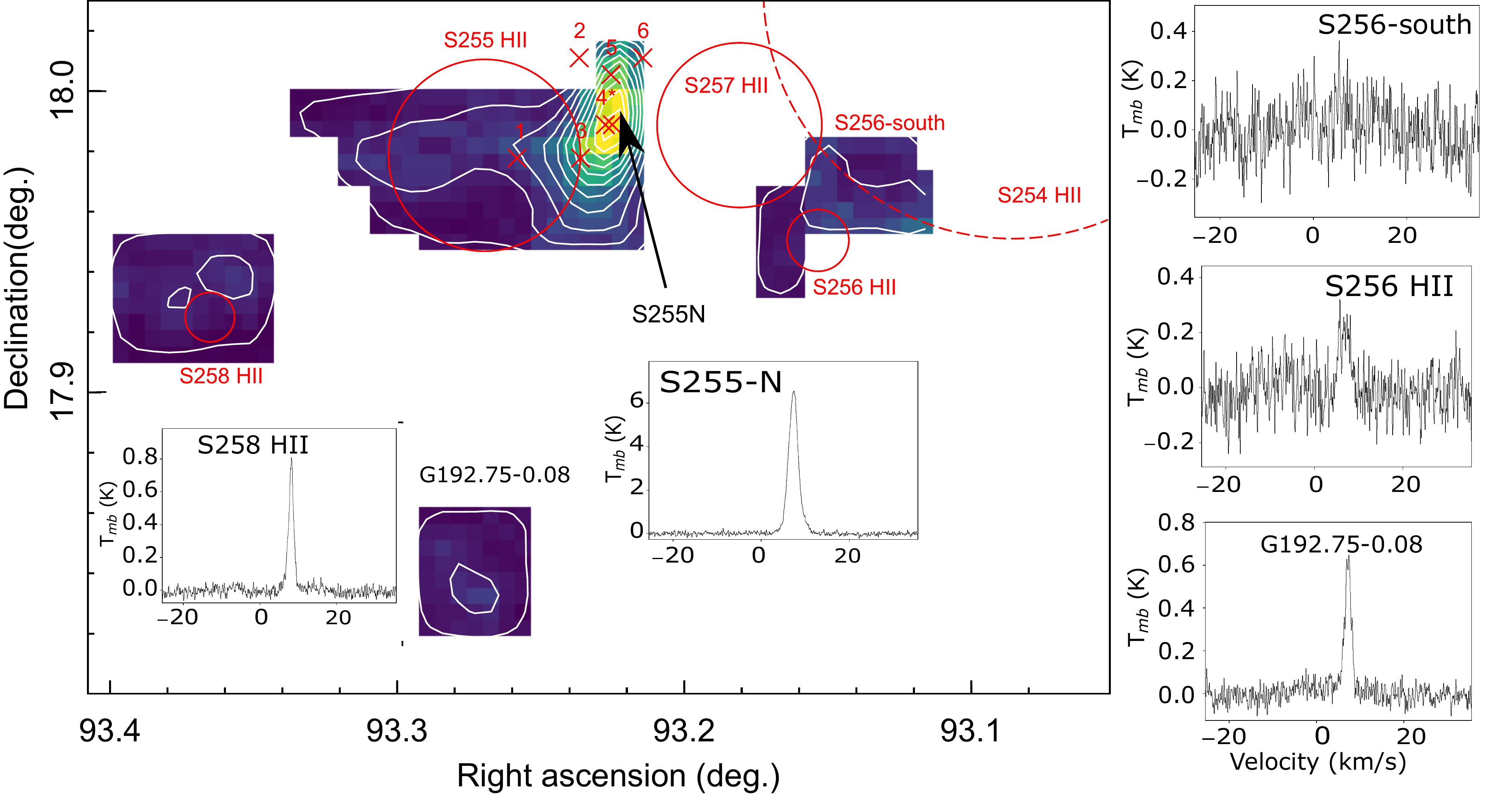}} 
\subcaptionbox{HCO$^+$ (1--0) integrated intensity}{\includegraphics [width=0.80\linewidth] {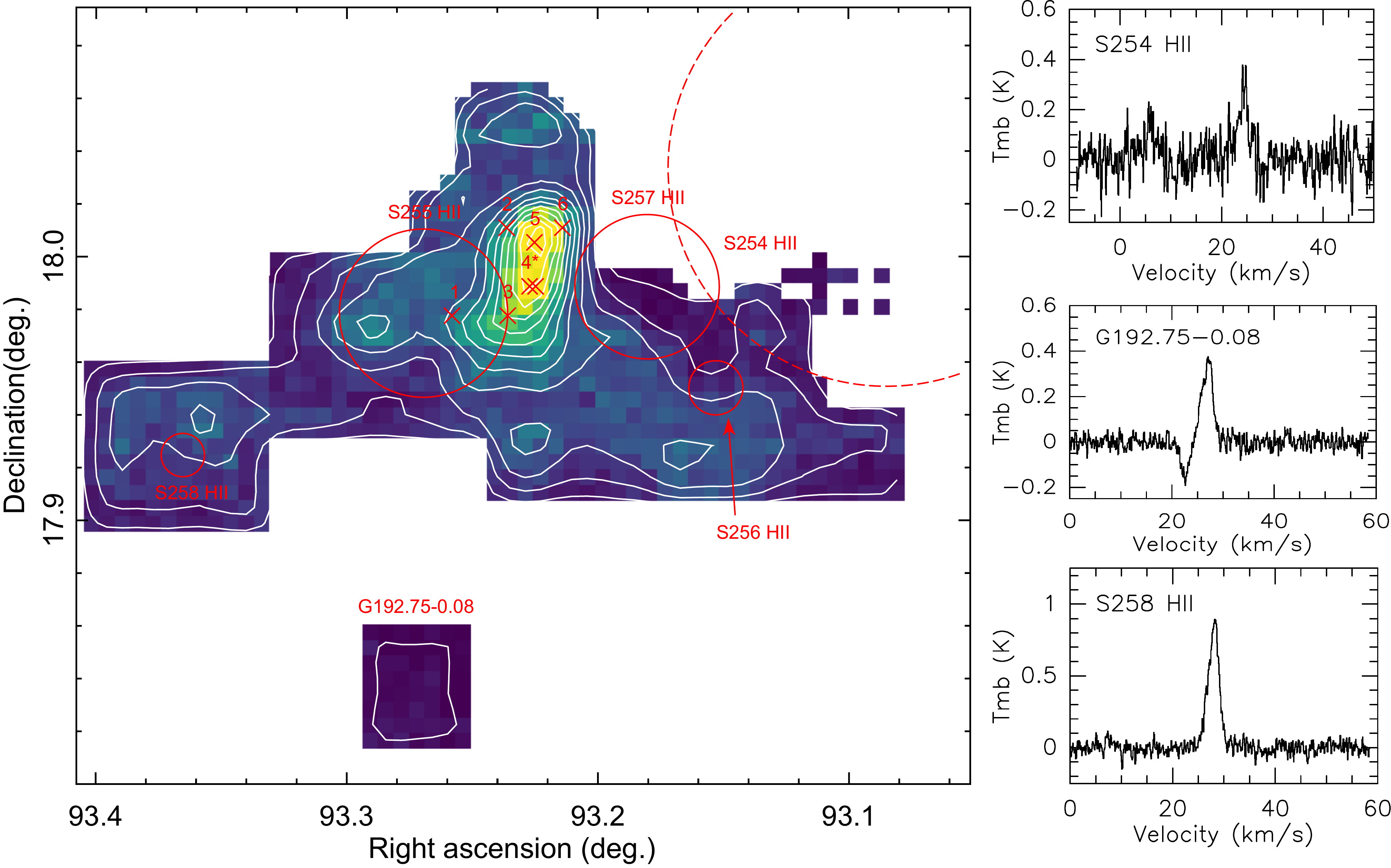}}

\caption [Maps of gas, dust and YSO] {Integrated intensity maps of the high density tracers in S254-258. Red circles represent visible radius of \hii{} regions from DSS red images. Spectra for several positions are shown in the sub-figures. (a) HCO$^+$ (1--0) map, contours go from 0.6 K~km~s$^{-1}$ to 5.6 K~km~s$^{-1}$ in steps of 0.45 K~km~s$^{-1}$. (b) CS(2--1) map, contours go from 0.56 K~km~s$^{-1}$ to 9.13 K~km~s$^{-1}$ in steps of 0.78 K~km~s$^{-1}$. 
}
\label{fig:maps_high_density}
\end{figure*}

We use the observations of \hco{}, CS and C$^{18}$O molecular line emission taken in February 2009 with the Onsala Space Observatory 20m radio telescope (PI -- Maria Kirsanova). The SIS receiver with the frequency range 85-116~MHz was used for the observations.  Extended mapping was done in CS(2--1) and HCO$^{+}$(1--0) lines toward the molecular ridge with a size of about $6$\arcmin$\times15$\arcmin. Fig.~\ref{fig:maps_high_density} shows the maps of the \hcoi{} and CS(2--1) integrated emission together with spectra in several selected directions.   Additionally to these maps we observed optically thin lines C$^{34}$S(2--1), C$^{18}$O(1--0), and H$^{13}$CO$^+$(1--0) toward ten selected directions, chosen to trace the inner and outer part of the central bright S255-IR cluster. These directions are marked by crosses in Figure~\ref{fig:maps_high_density}. Frequencies of the observed lines are given in Table~\ref{tbl_data_obs}. The total amount of the telescope time was 162~hours. Observations were taken with 40~MHz bandwidth and 1600 channels using frequency switching mode with a throw of 20~MHz. The spectral resolution corresponds to 0.07~km~s$^{-1}$. The full width at half-maximum (FWHM) of the telescope beam ranges from $34\arcsec$ to $43\arcsec$ and is listed in Table \ref{tbl_data_obs}. We used the beam size as spacing for the extended maps. 

\begin{table}
    \centering
 \caption{Characteristics of molecular line observations in the S255-258 region. $\theta_{\mathrm{FWHM}}$ is the beam size of the telescope at the specified frequency.}
  \label{tbl_data_obs}
 \begin{tabular}{@{}|l|c|c|c|c|c|c|c|c|}
\hline
Molecule & Line & Frequency &  $\theta_{\mathrm{FWHM}}$   & Observation type  \\
  &  &  (MHz) & (arcsec) &  \\
\hline
CS & (2--1)  & 97980.9 & 38.53 & Mapping  \\
\hco{} & (1--0)  & 89188.5 & 42.32 & Mapping   \\
C$^{34}$S & (2--1) & 96412.9 & 39.15 & Points   \\
C$^{18}$O & (1--0)  & 109782.1 & 34.38 & Points  \\
H$^{13}$CO$^+$ & (1--0) & 86754.2 & 43.51 & Points  \\
\hline
 \end{tabular}
\end{table}

The CLASS program from the GILDAS software package\footnote{\url{http://www.iram.fr/IRAMFR/GILDAS}} was used to process the obtained spectra. All spectra were converted to the T$_\mathrm{mb}$ scale using a primary beam efficiency. The beam efficiency value varies from 0.35 to 0.55, depending on the source elevation at the time of observation. We removed a 3rd order baseline from the raw data to obtain the continuum-subtracted spectra. The system temperature varies from 200 to 600~K, but in few cases exceeded 800~K. The spectra were averaged using the inverse system temperature as the weighting function. As we performed frequency-switching mode observations, we used the FOLD command of CLASS to deconvolve the frequency-switched spectra.

To estimate column density of CS and HCO$^+$ molecules, their excitation temperature and optical depth needs to be known. This requires the knowledge of at least two lines. In our case, we have maps in HCO$^+$(1--0) and CS(2--1), but H$^{13}$CO$^+$(1--0) and C$^{34}$S(2--1) data are available only at six positions. Therefore, we can only calculate the column densities in the directions of the available C$^{34}$S(2--1) and H$^{13}$CO$^+$(1--0) spectra. 

We obtained the line peak optical depth values ($\tau_0$) using the ratios of the peak main-beam temperatures of \hcoi{} and  CS(2--1) lines and corresponding isotopes ($\rmn{H^{13}CO^+}$(1--0), C$^{34}$S(2--1)). The same equation was used for calculation of the line-integrated optical depths ($\int\tau$) using the integrated main-beam temperatures:
\begin{equation}
\frac{T_\rmn{mb[isotope]}}{T_\rmn{mb[main]}}=\frac{1-e^{-\tau_0/R}}{1-e^{-\tau_0}};\,\, \frac{\int T_\rmn{mb[isotope]}}{\int T_\rmn{mb[main]}}=\frac{1-e^{-\int\tau/R}}{1-e^{-\int\tau}},
\label{eq:ratio}
\end{equation}

\noindent where the isotopic ratio $R$ is assumed $R(^{34}\mathrm{S}/^{32}\mathrm{S}) = 22$ and $R(^{12}\mathrm{C}/^{13}\mathrm{C}) = 80$. These values are consistent with the $^{12}$C/$^{13}$C ratio of 79.2 \citep{Milam05} toward S255 region at the galactocentric distance of 9.75 kpc.

The excitation temperature ($T_\rmn{ex}$) was estimated using the line peaks intensities and the optical depths ($\tau_0$) from the equation of radiative transfer:
\begin{equation}
T_\rmn{mb}=(J_v(T_\rmn{ex})-J_v(T_\rmn{bg}))(1-\exp(-\tau_0))
\label{Tex}
\end{equation}
\noindent where $J_v(T)=T_0/(\exp(T_0/T)-1)$ and $T_\rmn{bg}=2.7~$K using the coefficient $T_0=h\nu/k$. The used rest frequencies for different lines are listed in the Table~\ref{tbl_data_obs}.

The column densities of CS and HCO$^+$ were then calculated using the approach described by \citet{Mangum2015}. 
\begin{align}
N=\frac{\int\tau}{1-\exp(\int\tau)} \frac{\rm 3h}{8\pi^3S\mu^2}\,\frac{Q_\mathrm{rot}}{g_\mathrm{J}}\,\frac{\exp(\frac{E_{\rm u}}{{\rm k}T_\mathrm{ex}}))}{\exp(\frac{\rm h\nu}{{\rm k}/T_\mathrm{ex}})-1}\, \frac{1}{f}\int\tau\rmn{d}v
\label{eq:nthin}
\end{align}
where $S=J^2/(J[2J+1])$, $g_J=2J+1$ is statistical weight, $Q_\mathrm{rot}\simeq {\rm k}T_\mathrm{ex}/{\rm h}/B_\mathrm{rot} +1/3$ is partition function for linear molecules. We assume a filling factor $f=1$. The constants $B_\mathrm{rot}$, $E_\mathrm{u}$, and $\mu$ were taken form the Cologne Database for Molecular Spectroscopy \citep{Muller01} and given in Table~\ref{tbl_constants}.

\begin{table}
    \centering
 \caption{Constants that was used for column density calculation from the molecular line observations.}
  \label{tbl_constants}
 \begin{tabular}{@{}lccccccc}
\hline\hline
 Line & T$_0$ & B$_\mathrm{rot}$ & E$_\mathrm{u}$ & $\mu$ \\
      & K     & GHz              & K              & 10$^{-18}$ esu   \\
\hline
CO(1--0) &  5.532 & 57.6 & 5.53 &  0.11 \\
CO(2--1) &  11.064 & 57.6  & 16.60  & 0.11  \\
\coi{} &  5.288 & 55.1 & 5.29 &  0.11 \\
\coii{} &  10.577 & 55.1 & 15.87   &  0.11 \\
CS(2--1) &  4.702 & 24.5 & 7.10 & 1.96 \\
C$^{34}$S(2--1) &  4.627 & 24.1 & 6.3 &1.96  \\
\hcoi{} & 4.280  & 44.6 & 4.28 & 3.90  \\
H$^{13}$CO$^+$(1--0) &  4.163 & 43.3 & 4.16 & 3.90 \\
\hline
 \end{tabular}
\end{table}

\subsection{CO data}

We used \coi{} and \coii{}  data to study the whole molecular cloud associated with S254-258. The estimation of CO column density was done using the same technique that is described in the Section~\ref{hight_density_molecules}. For CO molecule we have $S=J^2/([2J+1])$, as described in the \citet{Mangum2015}. The data on $^{12}$CO(1--0) and $^{13}$CO(1--0) line emission is taken from \cite{Chavarria08}, data on $^{12}$CO(2--1) and $^{13}$CO(2--1) lines from \cite{Bieging09}. The optical depth of CO lines (both peak and integrated) were estimated from $^{12}$CO/$^{13}$CO line ratio (peak and integrated) using equation (\ref{eq:ratio}) and the abundance ratio $R=80$.  

The excitation temperature is determined from optically thick $^{12}$CO line. For $\tau\gg1$ the radiative transfer equation (\ref{Tex}) can be inverted to provide
\begin{equation}
\begin{aligned}
T_{\rm ex} = T_0 / \ln \left(1+\frac{T_0}{T_{\rm B}^{12}+T_0/(e^{T_0/T_\mathrm{bg}}-1)}
\right)
\end{aligned}
\label{A1}
\end{equation}
\noindent where $T_{\rm B}^{12}$ is the peak $^{12}$CO brightness temperature. 

We considered the use of gas and dust clump-extraction algorithms like \textsc{clumpfind} and \textsc{gaussclump} with available CO data  \citep{Chavarria08,Bieging09}. However, they would be limited by the relatively large velocity resolution of the available data ($1.2$~km~s$^{-1}$) that does not allow to distinguish kinematically different molecular clumps.

\subsubsection{Difference between CO(1-0) and CO(2-1) data}

Comparing the obtained values of line optical depth and excitation temperatures for the (1-0) and (2-1) transitions, we have found that excitation temperatures for both transitions are quite close (see Table~\ref{tbl:temp}), but the (2-1) transition has systematically lower excitation temperatures in comparison to the (1-0) transition. That can be due to radiation transfer effects as explained in details in \citet{Bernes79} (see Figure 1 therein) -- the (2-1) transition  traces more outer layers of the molecular cloud, less of the volume. For example, in Figure 1 of \citet{Bernes79} the (2-1) excitation temperature drops from 19~K in the cloud center to 12~K in the outer layers, while the excitation temperature (1-0) drops only from 19~K to 18~K. Thus, the excitation temperature of the (1-0) transition is much more close to the average molecular cloud's kinetic temperature than the excitation temperature of the (2-1) transition. The obtained values of optical depth for (2-1) and (1-0) reveal that the optical depth of (2-1) transition is 1.5-3 times larger than (1-0) transition consistent with the fact that the (2-1) line rather traces the surface.
Due to lower optical depth of the (1-0) transition it allows for a more accurate determination of excitation temperature being less sensitive to radiation transfer effects. Therefore we use only the (1-0) transition to further study the molecular gas column densities.

We calculate the column density of CO molecules with Equation (\ref{eq:nthin}), where we use the excitation temperature and the line-integrated optical depth of the (1-0) line. The constants of $^{13}$CO molecule that were used for column density calculation are presented in the Table~\ref{tbl_constants}. We convert N(CO) to N(H$_2$) using abundance [CO]/[H$_2$]~=~$8\times10^{-5}$ \citep{Blake87}. The distance to the S254-S258 region is assumed $1.59\pm0.07$ kpc \citep{Rygl10}, as measured by trigonometric parallax of 6.7 GHz methanol maser in the S255 region. We used this distance for the mass estimates in the current and subsequent sections.

\subsection{Extinction maps}\label{sec:Extinction_maps}

Extinction maps were produced using the NICEST method described by \citet{Lombardi09}. We used the implementation of \citet{Juvela2016}.

For constructing the extinction maps we used the data from the UKIRT Infrared Deep Sky Survey \citep[UKIDSS,][]{Lawrence07} together with data from the 2MASS survey \citep{Skrutskie06}.  The UKIDSS survey instrument is WFCAM \citep{Casali07} mounted on the UK Infrared Telescope (UKIRT) in Hawaii. The data from the WFCAM camera have a high quality -- resolution of 0.4~arcsec, photometric accuracy of $\sigma_J<0.02^m$ and a deep limiting magnitude up to $19.9^m$ in the $J$ band. The S254-S258 star formation region lies in the Galactic Plane Survey (GPS) area of UKIDSS. We constructed the extinction map using combined UKIDSS and 2MASS data centered on the S255N region with a radius of 1.5 degrees.

Due to the inaccuracy of the UKIDSS catalog for bright stars, we replaced the UKIDSS photometry by 2MASS for stars brighter than 12$^m$ for each band using the TOPCAT software \citep{Taylor05}. Photometry of 2MASS was corrected to the UKIDSS photometric standard using equations (6-8) from \citet{Hodgkin09}. 

We filtered the UKIDSS catalogue using the following criteria: 
\begin{enumerate}
\item Included sources are only those with a high probability of being a star (column ``\textsc{pstar}'' more than $0.8$ from the WFCAM catalogue);
\item At least one combination of two colours (J-H, J-K or H-K) available for each source; 
\item Sources that did not have at least two bands with magnitude uncertainties below 1$^m$ were rejected; 
\end{enumerate}

The NICEST method is based on the comparison between observed colours between the J, H and K bands and intrinsic colours of stars, corresponding to zero extinction. 

The NICEST maps were obtained at the resolutions (FWHM) of 1.0-3.0\arcmin. The pixel size was set to FWHM/3.  We used an extinction curve with $E(J-H)/A_V=0.10167$ and $E(H-K)/A_V=0.06347$ corresponding to the standard $R_V=3.1$ extinction curve \citep{Cardelli89} with $A_J/A_V=0.26$. We convert the extinction map to the gas column density using the approach described in the Appendix~\ref{app_extinction_h2}. The used conversion factor is $N(\mathrm{H})/A_V =1.87\pm0.13\times10^{21}~(\mathrm{cm}^{-2}\mathrm{mag^{-1})}$. The NICEST A$_J$ extinction maps at the best available resolution (1.0 arcmin) in the units of gas column density are presented in Fig.~\ref{fig:images_h2}.

The zero level of the NICER extinction map was set from a 
low-column-density region that is centered at RA$_\mathrm{J2000}=06^h15^m24.9^s$, Dec$_\mathrm{J2000}=18\degr33\arcmin45\arcsec$ and has a size of $40\arcmin\times20$\arcmin.

\subsection{Herschel data} \label{Herschel}

We study dust temperature and column density using the 70-500 \micron{} Hi-GAL data \citep{Molinari16}. We used two different data sets: the ViaLactea results \citep{Marsh17} based on the 70-500~\micron{} data and our own fit using only 160-500~\micron{} data. When comparing the ViaLactea total column density maps with the extinction maps we found that the values of the gas column density in the ViaLactea map is lower than the values from the extinction by a factor of 3--5, therefore we tried an independent fit. We believe that the discrepancy could be due to the missing Planck correction in ViaLactea map for the Herschel-PACS data making them inconsistent with the Herschel-SPIRE data in particular for the shortest wavelengths where the PLANCK data cannot be easily extrapolated. The detailed correction method for Herschel-PACS data using the Planck data can be found in \citet{Lombardi2014,Abreu-Vicente2017} for large-scale maps and in \citet{Sadavoy2018} for small-scale maps. However, in this work  we excluded the 70-100 \micron{} Herschel flux and performed a grey-body SED fit with a dust emissivity $\kappa \propto \lambda^{-1.8}$ only to the 160-500 \micron{} range. The resulting fit gives column densities and masses of the clumps that agree with CO-based values and the extinction maps, however it is not sensitive to warm dust ($>35$K). 

Therefore we used the ViaLactea results to create the contour levels for clumps and study the dust temperature of clumps, while the 160-500~\micron{} fit was used to estimate the column density and masses of the clumps and analyze the correlations between dust and YSOs density maps.

Using the ViaLactea map instead of the 160-500~\micron{} map gives almost no difference in the correlation analysis (Sections~\ref{correlation_power_law} and \ref{correlation_wwcc}): the correlation coefficients differ by less than 2\% between these maps. It only affects on the absolute values of the masses, column densities (Section~\ref{sec:mass_clumps}), and temperatures (Section~\ref{temp_dist}). We present both ViaLactea and 160-500~\micron{} fit results in the corresponding sections.

\subsection{Masses and error estimation}
We estimate each clump's mass using the integration of the clump area in the mass-in-pixel maps. The column density maps from each of the gas and dust tracers were converted to the maps of the mass-in-pixel using the following equation:
\begin{align}
\label{eq:mass_in_pixel}
M = N(\mathrm{H_2})\mu_{\mathrm{H}_2}m_\mathrm{H} \mathrm{A}
\end{align}
\noindent where $\mu_{\mathrm{H}_2}=2.8$ is the mean molecular weight of the interstellar medium \citep{Kauffmann08}, $m_H$ is the mass of an hydrogen atom, $\rm A$ is the area of all pixels of a clump in cm$^2$.

The errors on masses, column densities and temperatures are given in the Tables~\ref{tbl:mass_estimate}-\ref{tbl:temp}. They were estimated using \texttt{uncertainties} package of Python and root mean square (RMS) maps of the following tracers: $^{12}$CO(1-0), $^{13}$CO(1-0), Herschel (ViaLactea) 70-500 \micron, Herschel 160-500 \micron, and A$_J$ extinction. The dust temperature uncertainty map for ViaLactea \citep{Marsh17} is not available. Thus we do not include the errors in the Herschel 70-500 dust temperature estimation.

\subsection{Young stellar objects density maps} \label{YSO_map}

Using catalogs from \citet{Chavarria08}, we constructed density maps of class I and class II YSOs in order to study an evolutionary link between gas, dust and formation of YSOs. We created the map of the YSO density using the Kernel Density Estimate \citep[KDE;][]{Scott2015} with a Gaussian kernel, implemented in the KDEpy\footnote{\href{https://kdepy.readthedocs.io/en/latest/index.html}{https://kdepy.readthedocs.io/en/latest/index.html}} package of Python to make a direct comparison between the YSO density map and different gas and dust tracers. We tested different kernel widths in the range 45-180\arcsec{}. The kernel width was selected as 60\arcsec{} as a balanced value between smoothness and discretization. The resulting YSO densities are shown in Fig.~\ref{fig:yso}. The Wavelet-based cross-correlation (WWCC) analysis  \citep{Arshakian2016} provides a more detailed method of correlation analysis between the YSO density and gas and dust tracers (CO, HCO$^+$, extinction and Herschel maps) on different spatial scales. 

Before executing the correlation analysis, we convolved and regrid the studied maps to a uniform beam and uniform step size. The beam size of 60\arcsec{} matches the YSO density kernel size. After the convolution, the telescope beam has no impact on the correlation analysis, and one can compare all scales above that beam limit. The convolution and regridding were done using Python packages \textsc{astropy.convolution} and \textsc{reproject}, respectively. 

\begin{figure*} 
\vspace{5mm}
 \begin{minipage}{180mm}
 \center
\includegraphics [scale=0.88] {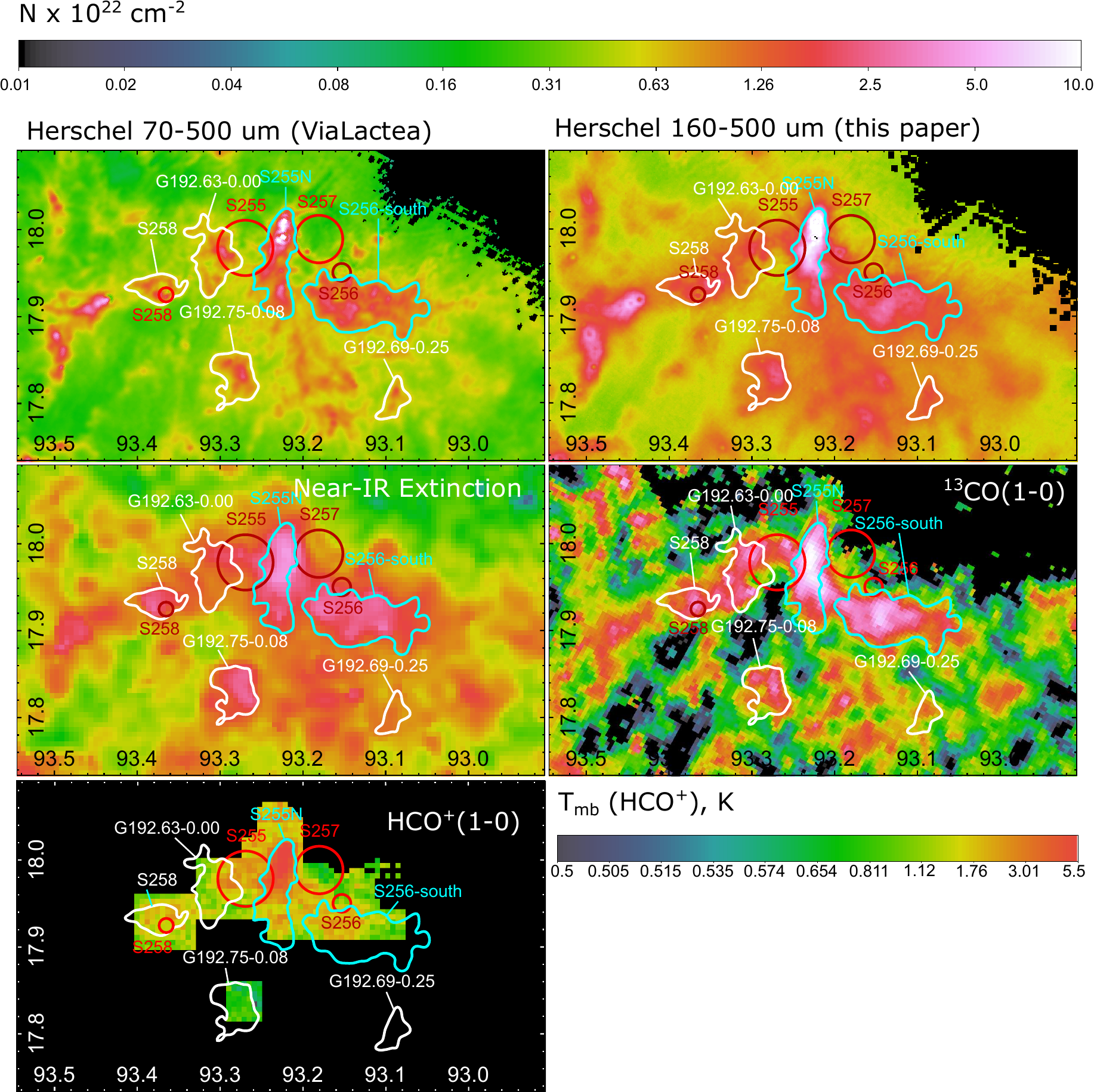}
 \caption {Panels 1-5: Maps of hydrogen column density obtained from different tracers of gas and dust in the S254-S258 region, including two Herschel maps from ViaLactea database \citep{Marsh17} and from this paper, \coi{} map \citep{Chavarria08}, Near-IR extinction map (see Section~\ref{sec:Extinction_maps} for details). In the last panel, the \hcoi{} intensity map is shown from the Onsala observations (this paper). All maps except \hcoi{} have the same units and scale of column density starting from 1$\times10^{20}$~cm$^{-2}$ and goes up to 1$\times10^{23}$~cm$^{-2}$. For the \hcoi{} map the intensity scale is shown in the lower right corner. In each panel, molecular clumps associated with star clusters are shown with a white and cyan thick line. The edges of molecular clumps marked with white borders are ViaLactea column density of $5.8\times10^{21}$~cm$^{-2}$ and cyan border are ViaLactea column density of  8.0$\times10^{21}$~cm$^{-2}$.  Red circles in each panel represent the visible radius of \hii{} regions from DSS-R survey. }
 \label{fig:images_h2}
\end{minipage}
\vspace{5mm}
\end{figure*}

\section{Results} \label{sec3}

\subsection{Gas and dust clumps} 

The structure of the S254-S258 star formation region seen by Herschel (see Figure~\ref{fig:images_h2}) maps reveals several bright gas and dust condensations, including S258, G192.63+0.00, S255N, S256-south, G192.75-0.08 and G192.69-0.25. 

We investigate only those molecular clumps that are associated with the star clusters from \citet{Chavarria08} (see Figure 12 therein). The borders of clumps were defined based on the ViaLactea column density map using thresholds of $5.8\times10^{21}$~cm$^{-2}$ for S258, G192.63-0.00, G192.75-0.08, G192.69-0.25 and of $8.0\times10^{21}$~cm$^{-2}$ for the S255N and S256-south regions. We used the two different thresholds for clusters to make the area of bright clumps associated with YSOs clusters (S255N, S256-south) similar to other clumps. Otherwise, the threshold of $5.8\times10^{21}$~cm$^{-2}$ will occupy the significant fraction of the S254-S258 star formation region and makes it impossible to distinguish between the gas related to YSOs clusters S255N and S256-south \citep{Chavarria08}. Due to the absence of bright dust emission in the Herschel bands in G192.75-0.00 and G192.54-0.15 clusters, we used contours of a YSO density at five stars per square parsec from \citet{Chavarria08} as the border for these clusters.

In the original work of \citet{Chavarria08} authors identify two YSOs clusters in the direction between S255 and S257 \hii{} regions: S255N\&S255-2 and G192.65-0.08. However, our multi-wavelength analysis (see Figure~\ref{fig:images_h2}) displays no difference between these clusters in any of the considered gas-dust tracers. Thus in further analysis we combine these YSOs clusters into a single cluster named S255N. 

The cluster G192.55-0.01  located to the north of the S255N region and contains only class II YSOs, but displays no emission in any of the available gas and dust tracers. Moreover, there is no pronounced border between the cluster and its surrounding class II YSOs. Thus we did not include it to our list.

The clusters S256-south, G192.63-0.00 and S258 are distinguished from the central region S255N in the Herschel map. Thus we consider them independently. However, CO and extinction maps display a smooth distribution of gas and dust toward S258, G192.63-0.00, S255N and S256-south. Thus these clusters may be considered as different parts of the main gas ridge in the S254-S258 star formation region. 

Clusters G192.75-0.08 and G192.69-0.25 have both class I and II YSOs and have pronounced peaks of emission in Herschel and CO maps. Thus they were included in our list.  

Cluster G192.75-0.00 contains only class II YSOs and has no pronounced emission in the Herschel map. However, it has pronounced peaks in the CO and extinction maps. Thus we include it to our list, but use the YSO density contour for the mass estimation instead of Herschel column density contours. 

\begin{table*}
\caption{Parameters of the observed lines at the different points obtained from the Gaussian fitting and physical conditions in the gas. The pointing center have coordinates $\alpha_\rmn{J2000}=6^{\rmn{h}} 13^{\rmn{m}} 4\fs1$, $\delta_\rmn{J2000} = 17\degr 58\arcmin 44\arcsec$. Values in the brackets are the formal errors from the \textsc{class} program. Point with asterisks mark (*) have offset of 8.7\arcsec{} in Right Ascension in different lines. $\tau_0$ is the optical depth at the line center channel, $\int\tau\rmn{d}v$ is line-integrated optical depth, $\rmn{T_{ex}}$ is the excitation temperature and $N$ is molecular column density. } 
\label{tbl:line_parameters}
\CSVtotabular{csv/fit.csv}{lcccccccccccc}{
\hline\hline
Line & Point &$\Delta\alpha$ & $\Delta\delta$  & $\rmn{V_{LSR}} $& $\rmn{\Delta V}$ &$\rmn{T_{mb}^{peak}}$ & $\int\rmn{T_{mb}}$ &  $\tau_0$ & $\int\tau\rmn{d}v$ & $\rmn{T_{ex}}$  & $N$ \\
& &  arcsec & arcsec & km\,s$^{-1}$ & km\,s$^{-1}$ & K &K km s$^{-1}$  &  & & K & 10$^{14}$~cm$^{-2}$ \\\hline}
{\insertLine &\insertPoint &\insertOffsetx &\insertOffsety &\insertVlsr &\insertdV &\insertTpeak &\insertArea & \insertTau & \insertTauint & \insertTex & \insertNtota \\}

\end{table*}

\begin{table}
\setlength{\tabcolsep}{3pt}
\small
\caption{Masses and average column densities  estimation of molecular clumps in S254-S258 region from different gas tracers. The value after the $\pm$ symbol is the $1\sigma$ RMS of the parameter estimate. }
\label{tbl:mass_estimate}

\CSVtotabular{csv/Mass_estimate.csv}{lcccccccccccc}{
\hline\hline 
\multicolumn{4}{c}{Clump mass, M$_{\odot}$} \\
&
\multicolumn{2}{c}{Herschel} &
\multicolumn{1}{c}{$A_\mathrm{J}$} &
 \multicolumn{1}{c}{CO}  \\
Name & ViaLactea & 160-500 \micron{} & & (1--0)  \\ \hline}
{\insertbyname{Name} & 
\insertbyname{Vialact M} & 
\insertbyname{Herschel M} & 
\insertbyname{Extinction M} & 
\insertbyname{CO10 M} \\
}{}
\CSVtotabular{csv/Mass_estimate.csv}{lcccccccccccc}{
\hline\hline 
\multicolumn{4}{c}{N, $\times10^{21}$~cm$^{-2}$} \\
&
\multicolumn{2}{c}{Herschel} &
\multicolumn{1}{c}{$A_\mathrm{J}$} &
\multicolumn{1}{c}{CO}  \\
Name & ViaLactea & 160-500 \micron{} & & (1--0) \\ \hline}
{\insertbyname{Name} & 
\insertbyname{Vialact N} & 
\insertbyname{Herschel N} & 
\insertbyname{Extinction N} & 
\insertbyname{CO10 N} \\ 
}{}
\end{table}

\begin{figure}
\centering
\includegraphics [scale=0.65] {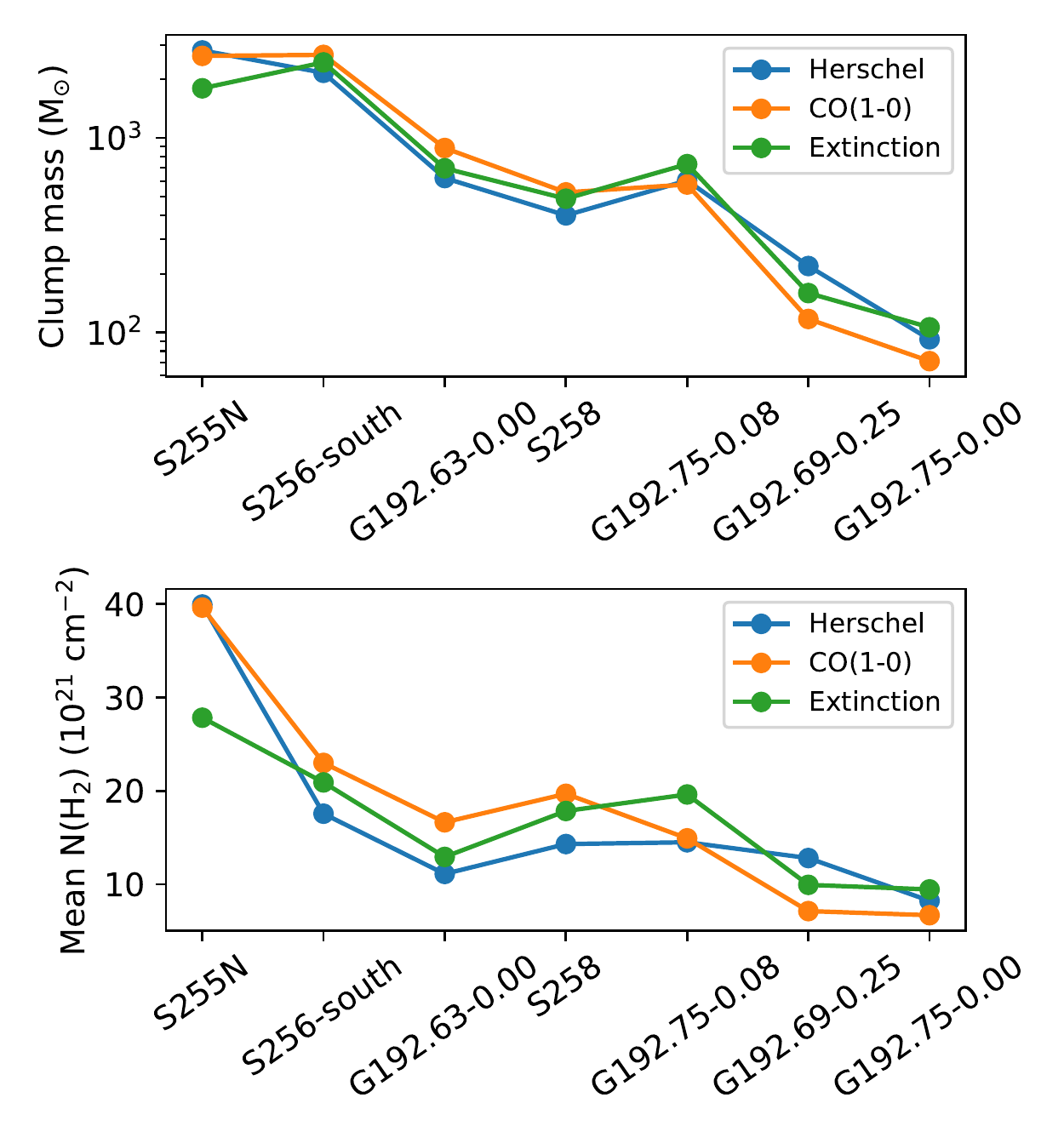}
\caption [Maps of gas, dust and YSOs] {Masses (upper panel) and average column densities (lower panel) of molecular clumps, associated with star clusters in S254-S258 region.
}
 \label{fig:mass_compare}
\vspace{5mm}
\end{figure}

\subsection{Masses and column densities of the clumps} \label{sec:mass_clumps}

\begin{figure}
\centering
\includegraphics [scale=0.65] {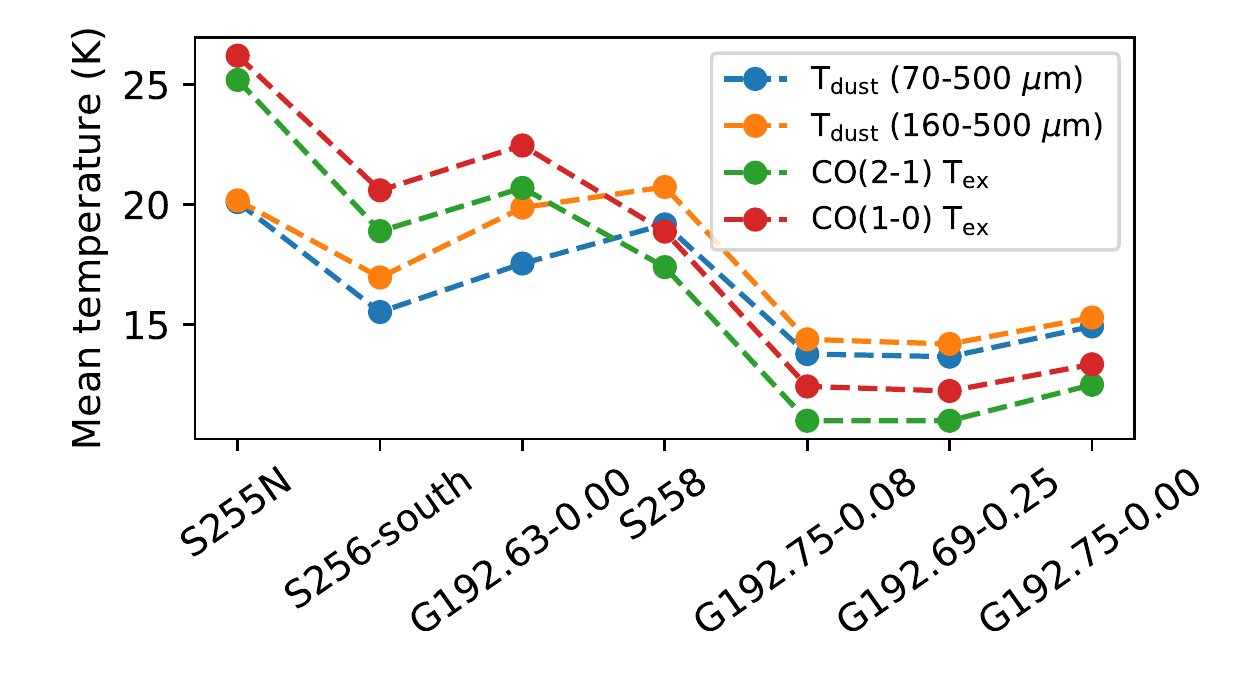}
\caption{Dust and CO excitation temperatures of molecular clumps, associated with star clusters in S254-S258 region. Designations are the same as in Table~\ref{tbl:temp}.
}
 \label{fig:temp}
\vspace{5mm}
\end{figure}

We divide all clusters into two main types: main molecular cloud and isolated clumps. Clusters S258, G192.63-0.00, S255N and S256-south have gas bridges in CO and the extinction maps. Thus they may be considered as different parts of the main molecular cloud in the S254-S258 region. Clusters G192.75-0.00, G192.69-0.25 and G192.75-0.00 are located at some distance to the main gas ridge and display no gas bridges with the main molecular cloud in any of available tracers (\coi{}, \coii{}, extinction, Herschel). Thus these clusters may be considered as isolated molecular clumps.

Table~\ref{tbl:mass_estimate} and Fig.~\ref{fig:mass_compare} present masses and column density estimates for the gas clumps, including errors. We found that, except for S255N, mass estimates from the CO, extinction and Herschel maps are close to each other.  The CO map gives the highest column-density and mass estimates in more massive clumps (S255N, S256-south, G192.63-0.00). 

The highest value of column density is found towards the S255N clump -- $N_\mathrm{H_2}=4.0\pm0.7\times10^{22}$~cm$^{-2}$ according to the CO and $4.0\pm0.2\times10^{22}$~cm$^{-2}$ according to the 160-500 \micron{} Herschel data. This value have the same order of column density of $\sim5.3\times10^{22}$~cm$^{-2}$ from the C$^{18}$O spectral line observations of \citet{Minier05}. Mass estimation for S255N clump is $2638\pm50$ $M_{\odot}$ according to CO and $2813\pm13$ $M_{\odot}$ according to Herschel data, however extinction map gives only $1797\pm21$ $M_{\odot}$. Undersampling of the extinction map and lack of background stars in S255N may cause this discrepancy.

The most massive clumps from all considered here are S255N and S256-south ($2638 \pm 49.8$ and $2669 \pm 50.7$ $M_\mathrm{\odot}$ respectively, according to \coi{} data). According to the Herschel map, they contain 71\% of the mass of our selected clumps and 27\% of the mass for the whole studied region (21\arcmin{}$\times$15\arcmin{}) containing all other clumps and inter-clump medium. In contrast, the area of these clumps is the only 12\% of the studied region.

\begin{table}
\setlength{\tabcolsep}{3pt}
\small
\caption{Dust and CO excitation temperature estimation of molecular clumps in the S254-S258 region. 70-500 Herschel T$_\mathrm{dust}$ values refer to the mean dust temperatures obtained using the 70-500 \micron{} fit from \citet{Marsh17}. 160-500 Herschel T$_\mathrm{dust}$ values were obtained from the fit described in Section~\ref{Herschel}. CO excitation temperatures for (1-0) transition were obtained using FCRAO data \citep{Chavarria08} and SMT data for (2-1) transition \citep{Bieging09}. The value after the $\pm$ symbol is the $1\sigma$ error of the parameter estimate.  }
\label{tbl:temp}
\CSVtotabular{csv/Mass_estimate.csv}{llllllllllllll}{
\hline\hline 
&
\multicolumn{2}{c}{T$_\mathrm{dust}$, K} &
\multicolumn{2}{c}{T$_\mathrm{ex}$, K} \\
&
\multicolumn{2}{c}{Herschel} &
\multicolumn{2}{c}{CO} \\
Name & 70-500  & 160-500 \micron{}  &  (1--0) & (2--1) \\ \hline}
{\insertbyname{Name} & 
\insertbyname{Tdust} & 
\insertbyname{Tdcorr} & 
\insertbyname{Tex10} & 
\insertbyname{Tex21} \\ 
}{\insertbyname{Name} & 
\insertbyname{Tdust} & 
\insertbyname{Tex10} & 
\insertbyname{Tex21} \\  \hline
}
\end{table}

\subsection{High-density gas}

Table~\ref{tbl:line_parameters} presents the observed parameters of  C$^{34}$S(2--1) and H$^{13}$CO$^+$(1--0) lines together with the CS(2--1), HCO$^+$(1--0) in similar offsets to compare the line intensities. . Results of the column density estimation are given in the Table~\ref{tbl:line_parameters}. Analyzing the values of CS column densities, we found them in the range $N_\mathrm{CS}=(0.10-6.31)\times10^{14}$~cm$^{-2}$ for the outer and inner parts of S255N cluster, respectively. Assuming relative abundance of CS to H$_2$ in the range 10$^{-9}$-10$^{-8}$ we estimate the maximum beam-averaged gas column density in the range $N_\rmn{H_2}=0.63-6.31\times10^{23}$~cm$^{-2}$. For comparison, \citet{Zinchenko09} estimated the gas column densities toward S255N as $N_\rmn{H_2}=2.0\times10^{23}$~cm$^{-2}$.

The available high-density tracer observations confirm the detection of \hcoi{} line toward S258, S256, S255, S257 \hii{} regions and G192.75-0.08, G192.63-0.00, S256-south star cluster. The CS(2-1) line is detected toward S258, S256, S255 \hii{} regions and G192.75-0.08, G192.63-0.00 star cluster. 

We confirm the presence of high-density gas traced by HCO$^+$ line emission in the inter-clump bridge between the S255N and S256-south star clusters.The typical intensity of \hcoi{} line in the inter-clump bridge is 1.8~K, while $\sigma$ is 0.3~K, leading to a detection at 6$\sigma$ level. The presence of high-density gas in the inter-cluster medium suggest that these clusters are physically associated and share the same gas and dust material. We also searched for the inter-clump bridge between S255 and S258 \hii{} regions, but we detected no \hco{} emission at the level $\sigma \sim0.3~$K. That may lead to the conclusion that these \hii{} regions do not share the same gas and dust material and maybe are not physically associated. We also note that the $^{13}$CO column density map (see Figure~\ref{fig:images_h2}) reveals diffuse emission in the inter-clump medium between the S255 and S258 \hii{} regions. However, this emission may be due to the large extent of the gas along the line of sight that leads to high values of column density.

There are other meaningful differences between various maps in Fig.~\ref{fig:images_h2}. First, the peak of the S255N region in the \hco{}, Herschel maps is shifted to the north in comparison to the same peak in the $^{13}$CO and extinction maps. The shift may be related to a different distribution of high-density and intermediate-density gas inside the S255N cluster. 

We used the available observations of C$^{18}$O molecule in two points to check the optical depth of $^{13}$CO line used for mass estimations (see Section~\ref{sec:mass_clumps}). In Point~1 (see Table~\ref{tbl:line_parameters}) the peak of the C$^{18}$O(1--0) line is 1.12~K, while the $^{13}$CO(1--0) line has a main-beam temperature of 7.2~K. From the $^{12}$CO(1--0) and $^{13}$CO(1--0) line ratios and equation (\ref{eq:ratio}) we obtain $\tau_0=30$ for the $^{12}$CO(1-0) line. The abundance ratio $R= 80$ then gives $\tau_0=0.38$ for the $^{13}$CO(1--0) line. Using the ratio between $^{13}$CO(1--0) and C$^{18}$O(1--0) lines in the equation (\ref{eq:ratio}) we obtain the  ratio of optical depth for $^{13}$CO/C$^{18}$O $R=7.5$. The same calculations were done for Point~2, where we obtain $R=8.3$. The resulting values match the $^{13}$CO/C$^{18}$O abundance ratio R$\sim$7.3 found by \citet{Schoier02}.

\subsection{Temperature distribution} \label{temp_dist}

\begin{figure}
\centering
\includegraphics [scale=0.4] {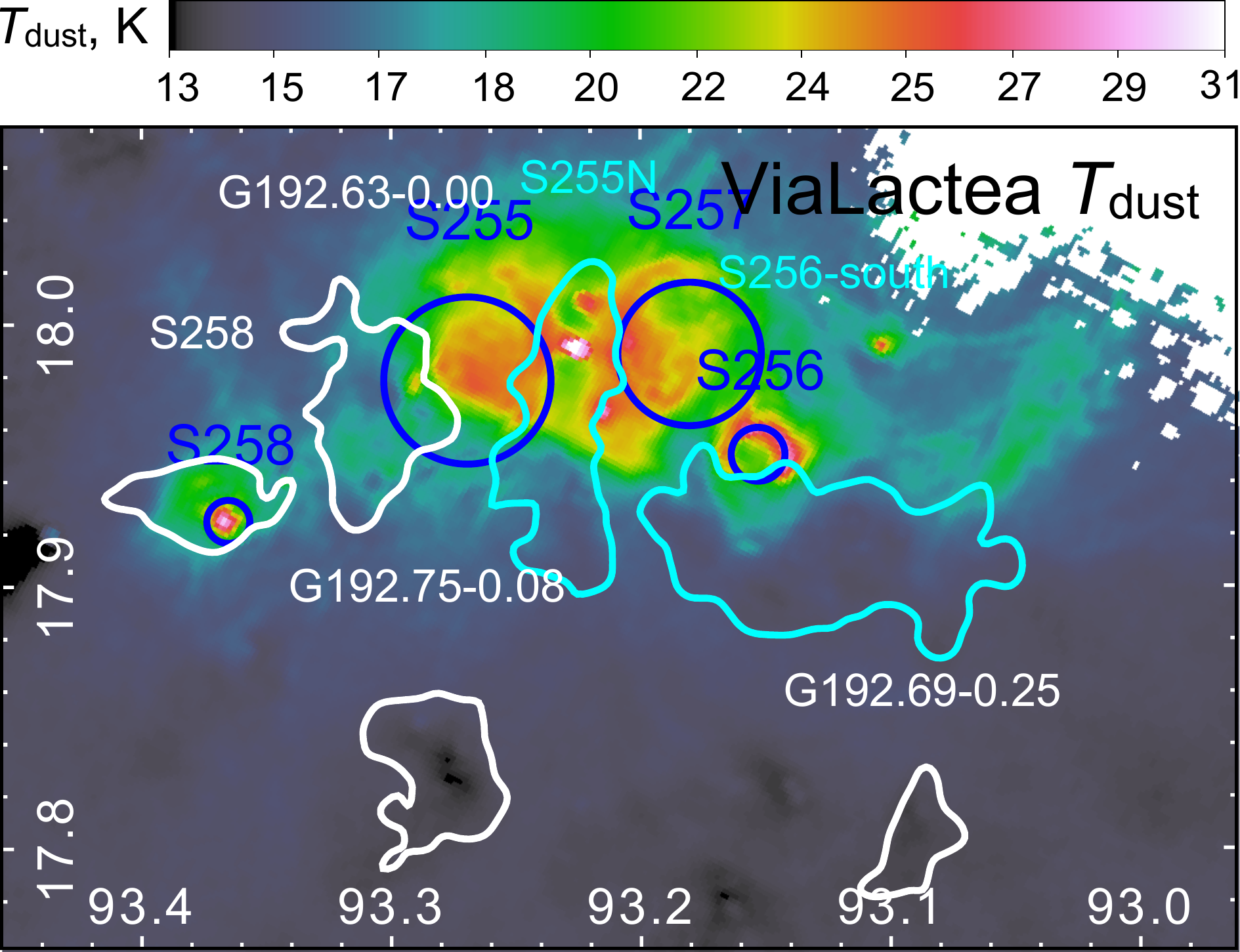}
\caption [Dust temperature] {ViaLactea dust temperature map of S254-S258 region. Contours are star clusters associated with high-density gas, same as in the Figure~\ref{fig:images_h2}. Blue circles are \hii{} regions, same as in the Figure~\ref{fig:images_h2}.
}
 \label{fig:tdust}
\vspace{5mm}
\end{figure}

The dust temperature estimation from the ViaLactea maps gives 17-20~K for S255N, G192.63-0.00, S258 and 14-16~K for S256-south, G192.75-0.08, G192.69-0.25 and G192.75-0.00. In the work of \citet{Zinchenko97} NH$_3$ molecular line observations of S255N region reveal gas kinetic temperatures 23.0~K and 24.7~K at the S255 core peaks. Their positions is (80\arcsec,0\arcsec) and (40\arcsec,--40\arcsec) in respect to the center field coordinates RA$_\mathrm{J2000}=06^h12^m53.3^s$, Dec$_\mathrm{J2000}=17\degr59\arcmin22\arcsec$. These positions are close to point~5 (distance $\approx$~25\arcsec) and 6 (distance $\approx$~8\arcsec) from the Table~\ref{tbl:line_parameters}. The temperatures drop to 15-20~K at the edges of the ammonia emitting regions. Therefore, gas and dust temperatures agree with each other in these dense molecular clumps.

The dust temperature ($T_{\rm d}$) obtained from the Herschel data are shown in Fig.~\ref{fig:tdust} and the mean values are presented in Table~\ref{tbl:temp} and Figure \ref{fig:temp} together with excitation temperatures of the \coi{} and \coii{} lines. We found that $T_{\rm d}$ is lower than the CO excitation temperature (both (2--1) and (1--0)) in more massive clumps ($>600M_\odot$). In contrast, the $T_{\rm d}$ value is higher than the CO excitation temperature (again both (2--1) and (1--0)) in less massive clumps.  We note that the more massive clumps are associated with the developed \hii{} regions, and the different behaviour can be explained by the influence of the \hii{} regions in the heating of the interstellar medium around these regions. Photon-dominated regions (PDR) around \hii{} regions always show a gas temperature exceeding the dust temperature close to the UV sources while the dust temperature exceeds the gas temperature at larger distances \citep[see e.g.][]{Roellig2013}.

We note that the mean the dust temperature of S256-south (15.5~K) and S258 regions (19.2) are significantly different (see Figure~\ref{fig:tdust} and Table~\ref{tbl:temp}), while the mean \hcoi{} line intensity is almost similar for both of them -- 1.6~K for S258 and S256-south regions (see Figure~\ref{fig:images_h2}). The physical interpretation is that the dust temperature map measures total energy input from heating sources, while \hco{} emission shows PDR  material, i.e., traces the heating sources' UV radiation. We conclude that the heating source in the S258 \hii{} region produces more thermal emission than in the S256 \hii{} region. According to \citet{Russeil2007}, the spectral type of the heating source is B3V for the S258 and B2.5V for the S256. Although the difference in spectral class is minor, the difference in the radiation spectrum is significant. The radiation fields' difference may be associated with multiple heating sources in the S258 \hii{} region. The YSOs cluster that ignites the S258 \hii{} region may harbor several stars that do not emit intense UV radiation and heat the surrounding gas. This radiation may result in a higher dust temperature in the S258 region than in the S256-south region. However, more observational data are needed on the S258 and the S256-south region to clarify the situation.

\begin{figure*}
\centering
\includegraphics [width=0.8\linewidth] {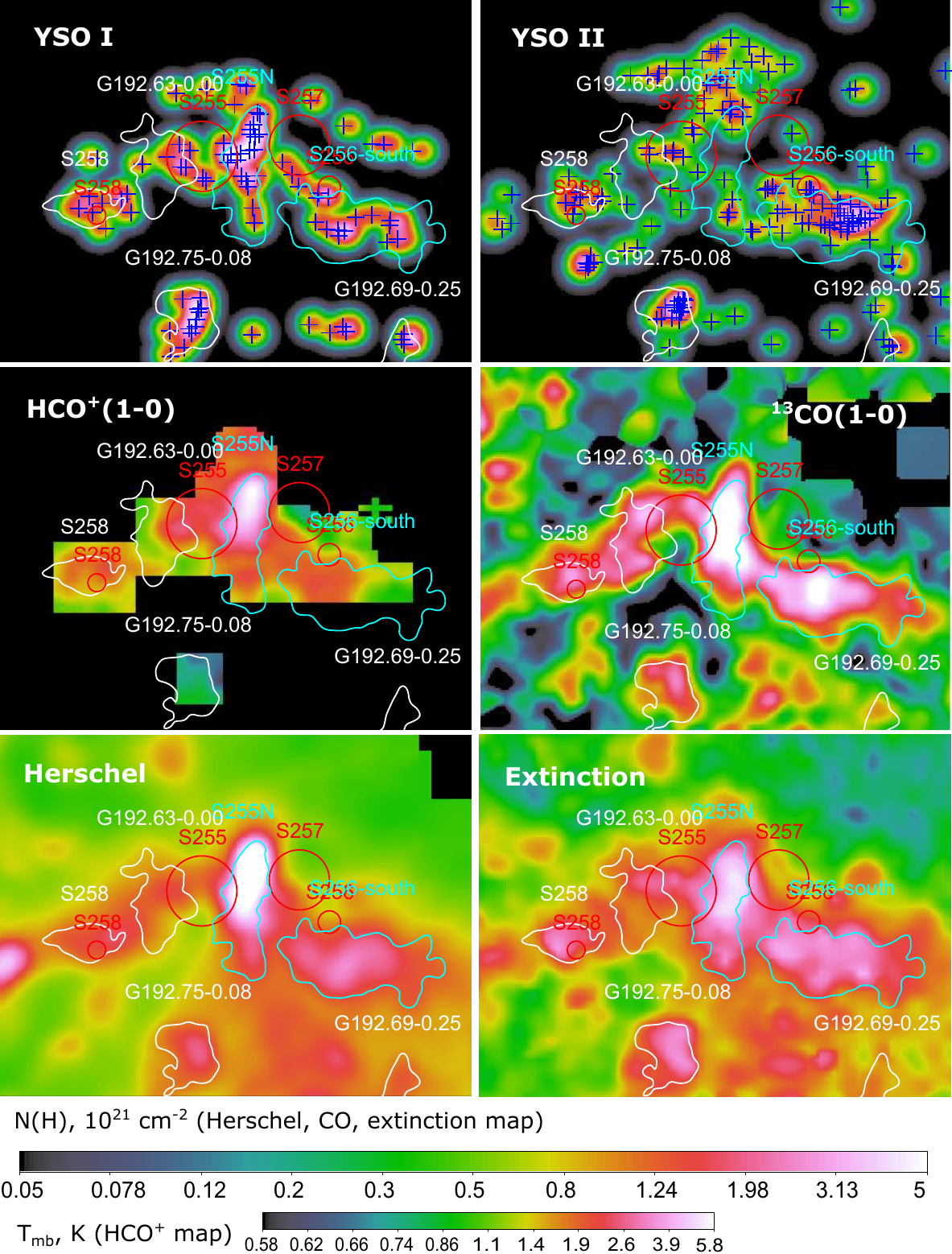}
\caption [YSO density map] {Input maps that were used for correlation analysis. YSOs density maps are named as YSO I and YSO II according to YSOs class and have arbitrary units. Herschel, \coi{} and extinction maps have units of column density and the same scale and limits, shown at the bottom right corner. Map of \hcoi{} line emission have T$_\mathrm{mb}$ units. All maps are log-scaled by the base of 100. White contours follow ViaLactea column density thresholds for selected YSOs clusters (same as in the Figure~\ref{fig:images_h2}). Red circles represent the visible radius of \hii{} regions. Blue crosses in YSOs density map mark the position of individual YSO.
}
 \label{fig:yso}
\vspace{5mm}
\end{figure*}

\begin{table*}
\caption{Correlation coefficients and power law indices (slopes) between YSOs class I and II density and different tracers of gas and dust convolved to the beam size of 60 arcsec. Values in the brackets are the 95\% confidence intervals of the correlation coefficient -- no error propagation was used here and the scatter is purely statistical. ``Without S255N'' region refers to the S254-S258 region not including the central bright S255N cluster (center at J2000 position 6:12:54.2, +17:58:56; size 3.75$\times$5.54 arcmin). The ``\hco{} mask'' region refers to the region masked with the available coverage map of \hco{} line -- see blue polygon in the Figure~\ref{fig:yso}. \label{tab_cc}}
\setlength{\tabcolsep}{3pt}
\small
 \begin{tabular}{llccccccccccc}
\hline\noalign{\smallskip}
    gas and dust tracer & Region & \multicolumn{3}{c}{Class I YSOs}   &  \multicolumn{3}{c}{Class II YSOs}  \\
     & & $r$ [$r_\mathrm{low}$, $r_\mathrm{high}$] & slope  &  $\chi^2$ & $r$ & slope  &  $\chi^2$ \\
  \hline
     Herschel 160-500 micron{} & Full &  0.75 [ 0.64,  0.83 ]  &  1.02 $\pm$  0.47 &  0.04  &  0.42 [ 0.29,  0.54 ]  &  0.31 $\pm$  0.24 &  0.05   \\ 
 & Without S255N &  0.47 [ 0.25,  0.64 ]  &  0.43 $\pm$  0.67 &  0.02  &  0.52 [ 0.40,  0.63 ]  &  0.33 $\pm$  0.24 &  0.03  \\ 
 & \hco{} mask &  0.69 [ 0.54,  0.80 ]  &  0.97 $\pm$  0.57 &  0.05  &  0.25 [ 0.07,  0.42 ]  &  0.18 $\pm$  0.36 &  0.03  \\ \hline
\coi{} & Full &  0.60 [ 0.44,  0.72 ]  &  1.44 $\pm$  0.49 &  0.19  &  0.41 [ 0.27,  0.53 ]  &  0.74 $\pm$  0.24 &  0.30  \\ 
 & Without S255N &  0.32 [ 0.08,  0.52 ]  &  0.87 $\pm$  0.70 &  0.22  &  0.44 [ 0.30,  0.56 ]  &  0.78 $\pm$  0.25 &  0.29  \\ 
 & \hco{} mask &  0.52 [ 0.31,  0.67 ]  &  1.17 $\pm$  0.57 &  0.18  &  0.22 [ 0.03,  0.39 ]  &  0.42 $\pm$  0.36 &  0.26  \\  \hline
A$_J$ extinction & Full &  0.69 [ 0.56,  0.79 ]  &  0.68 $\pm$  0.47 &  0.03  &  0.57 [ 0.46,  0.66 ]  &  0.51 $\pm$  0.24 &  0.06  \\ 
 & Without S255N &  0.52 [ 0.32,  0.68 ]  &  0.57 $\pm$  0.67 &  0.03  &  0.61 [ 0.50,  0.70 ]  &  0.54 $\pm$  0.24 &  0.05  \\ 
 & \hco{} mask &  0.64 [ 0.47,  0.76 ]  &  0.52 $\pm$  0.57 &  0.02  &  0.50 [ 0.34,  0.63 ]  &  0.29 $\pm$  0.36 &  0.02  \\  \hline
\hcoi{} & \hco{} mask &  0.44 [ 0.22,  0.61 ]  &  0.51 $\pm$  0.57 &  0.05  & -0.62 [-0.72, -0.48 ]  & -0.42 $\pm$  0.36 &  0.02  \\ 
  \hline
  {\smallskip}
\end{tabular}
\end{table*}

\section{Correlation between gas and YSOs}\label{sec:correlation_analysis}

Fig.~\ref{fig:yso} presents YSO density map for both class I and class II YSOs, created using KDE method (see details in Sec.~\ref{YSO_map}). From  visual inspection of Fig.~\ref{fig:yso} , one can notice that YSOs of both classes (I and II, see panels 1-2 on the Fig.~\ref{fig:yso}) are associated with the regions of high H$_2$ column density (see panels 4-6 on Fig.~\ref{fig:yso}). We used two independent methods to quantify differences in the distribution of the of class~I and II YSOs and their relation to various gas and dust tracers. The first one is the power law fit to the YSOs density vs gas column density, similar to one in \citet{Gutermuth2011}. The second one is correlation analysis on the different spatial scales using Fourier transformation with the WWCC method \cite{Arshakian2016}.  The following gas and dust tracers were used to study their relationships to YSOs density: extinction, \coi{}, Herschel 160-500~\micron{} and \hcoi{}.

\subsection{A power law fit to the pixel-to-pixel map of YSOs density and gas column density} \label{correlation_power_law}

The work of \citet{Gutermuth2011} reveals a power law dependence between YSOs density and gas column density. The power law is found in scatter plots of YSOs density against gas and dust column densities. The slopes are found to be 2.67 and 1.87, while the correlation coefficients are 0.87 and 0.83 for MonR2 and Ophiuchus molecular clouds, respectively. In the work of \citet{Bieging16}, those authors report a power law dependence of the mean YSO surface density on total gas surface density with the slope of 1.63.  We check these relations using the available gas and dust tracers of the S254-S258 region. The column density maps obtained in this paper for each tracer were used to plot the dependence of gas column density to the class I and II YSOs density in the direction of each YSO (similar to Figure 9 of \citet{Gutermuth2011}). 

Results of the power law fits and direct correlation analysis for class I and class II YSOs are presented in the Table~\ref{tab_cc}. We convolved each map to the beam size of 60\arcsec{} before going to calculations. The beam size was chosen as maximum beam size from the avaliable maps (extinction map). We also tested other beam sizes for calculations. The calculations for 90\arcsec{} beam size resulted in the relative increase (8-12\%) of correlation coefficient ($r$) in comparison to 60\arcsec{} calculations. For calculation of $r$ we used the log-based values of gas column density, main-beam temperature and YSOs density. The calculations were done for several cases:
\begin{enumerate}
    \item We calculated correlation coefficients and power law slopes for the whole studied region.
    \item We excluded the brightest part of S255N cluster to study the surrounding gas around the S255N region.
    \item We masked all studied maps using available \hcoi{} map coverage to compare the \hco{} map with other gas tracers.
\end{enumerate}

Analysis of the correlation coefficients between class I YSOs and various gas tracers in a whole studied region (i) reveal a high degree of correlation ($r=0.60-0.75$) with following maps: Herschel, \coi{} and extinction. The \hco{} map has a lower value of $r=0.44$. However, the lower value of $r$ may be related to the different unit scale of \hco{} map -- main-beam temperature instead of column density for other gas tracers. Using the direct values of main beam temperature instead of log-based values increase the value of $r$ from $0.44_{-0.22}^{+0.17}$ to $0.62_{-0.17}^{+0.13}$ for \hco{} map. 

Similar to \citet{Gutermuth2011}, we found a power law dependence between YSOs density and column density of various gas tracers.  The average slope of power law between YSOs of class I and various gas tracers is 0.9, while slope values range from 0.51 (\hco{} map) to 1.44 (CO map). The found values are generally lower than those found for MonR2 (2.67) and Ophiuchus clouds (1.87) in \citet{Gutermuth2011} and the slope of 1.63 for the S235 region found in \citet{Bieging16}.

We further studied the behaviour of correlation coefficients on masked regions. Exclusion of the central bright S255N region (ii) resulted in a change of correlation coefficients in Herschel,  and \coi{} maps -- the values are decreased from 0.6-0.7 to 0.3-0.4. We conclude that a high degree of correlation between those maps and class I YSOs is associated with the bright S255N region, rich in class I YSOs. Other regions in those maps display relatively high values of $r=0.5-0.6$. However, for the extinction map, the exclusion of the S255N region does not affect of correlation coefficient significantly - the value of $r$ is decreased from 0.69 to 0.52. We associate this with S256-south, S258 and G192.75-0.08 regions that all have high column densities ($N(\mathrm{H_2})\sim14-19\times10^{21}$~cm$^{-2}$, see Table~\ref{tbl:mass_estimate}) and rich in class I YSOs (see Figure~\ref{fig:yso}). 

We note that \hco{} map has a limited coverage, thus the results for \hco{} would be much stronger  and are subject to change if the mapping of the whole region becomes available. In order to compare the correlation coefficients of \hco{} map with other maps we have to apply the same \hco{} coverage mask (iii) to all other maps. The region with available \hco{} data is marked as ``\hco{} mask'' in the Table~\ref{tab_cc}. The masking resulted in decrease of correlation coefficient for all studied maps from 0.6-0.75 to 0.52-0.69. Thus we conclude that correlation coefficient between \hco{} map and class I YSOs density map is decreased due to limited coverage of the \hco{} map. This is also confirmed in the spatial correlation analysis in the ``HCO-mask'' region (see Section~\ref{correlation_wwcc}), where values of $r$ are almost similar for all gas and dust tracers within the ``HCO-mask'' region.  

The correlation coefficients and power law slopes between class II YSOs and different gas and dust tracers are lower than for the class I YSOs. The values of $r$ range from 0.41 (CO map) to 0.57 (extinction map). The slope for power law is  shallower than the similar slope for class I YSOs and have a mean value of 0.52 for those tracers that have $r>0$.  The exclusion of the S255N region results  in an increase in correlation coefficients from 0.29-0.57 to 0.38-0.61. We associate the increase with the small amount of class II YSOs in the S255N cluster (see Figure~\ref{fig:yso}), but high values of column density toward this cluster (see Table~\ref{tbl:mass_estimate}), the maximum for the whole S254-S258 region. Excluding such a region removes the main difference between maps of YSOs class II and gas tracers.

Another notable feature is the negative value of $r=-0.62$ for \hco{} map and class II YSOs density. If one uses the direct values of \hcoi{} main-beam temperature instead of log-based values, this results in $r=-0.50^{+0.15}_{-0.13}$. Thus the HCO$^+$ map reveals a statistically significant ($p<0.001$) negative value of correlation coefficient $r=-0.58^{+0.14}_{-0.12}$ with class II YSOs density. Visual inspection of HCO$^+$ and class II YSOs density maps confirms the found negative relation: the peaks of \hco{} emission do not coincide with maxima in class II YSO density (see Figure~\ref{fig:yso}). 

In the S255N region the peak of YSOs class II density has an offset of $\sim$3 arcmin to the north and to the south from the \hco{} emission peak. In the S256-south region the peak of class II YSOs density have an offset of $\sim$2.5 arcmin to the east from the \hco{} emission peak. Only the S258 region displays no offset between YSOs class II and \hco{} emission peak. The G192.75-0.08 cluster is covered by line observations but is not associated with strong \hco{}  emission. An average main-beam temperature of HCO$^+$ line in G192.75-0.08 is 0.7 K, while in the S258 and S255N regions it has values of $\sim2$~K and $\sim5$~K, respectively.

We conclude that class II YSOs density is anti-correlated to the high-density gas traced by \hco{} map. At the same time, the class II YSOs density map reveals a weak positive correlation with other tracers of gas and dust including extinction, \coi{} and Herschel maps with the value of $r=0.41-0.57$. In contrast, class I YSOs have much more pronounced positive correlation with all studied maps with the values of $r$ range from 0.60 to 0.75. 

\subsection{Correlation on different spatial lengths} \label{correlation_wwcc}

Direct comparison between YSOs density and gas column density does not give information about the spatial scale at which correlation becomes more pronounced. To reveal this spatial scale and measure the dependence of correlation coefficients on spatial scale quantitatively, we use the WWCC method \citep{Arshakian2016}. This method gives the possibility to find correlation coefficients between two images on different scales using Fourier decomposition of input maps to the different scales and subsequent calculations of correlation coefficients on different scales. For calculations, we use three cases, similar to those described in the Section~\ref{correlation_power_law}: the whole S254-258 region, the region without the S255N cluster and the region masked with the \hco{} map coverage. Excluding the S255N region allows one to study the surrounding gas around the main cluster. Masking all tracers by the \hco{} map coverage allow us to compare the \hcoi{} map with all other tracers. 

\begin{figure*} 
\vspace{5mm}
 \begin{minipage}{180mm}
 \center
\subcaptionbox{Class I YSOs (Whole region)}{\includegraphics[width=0.47\textwidth]{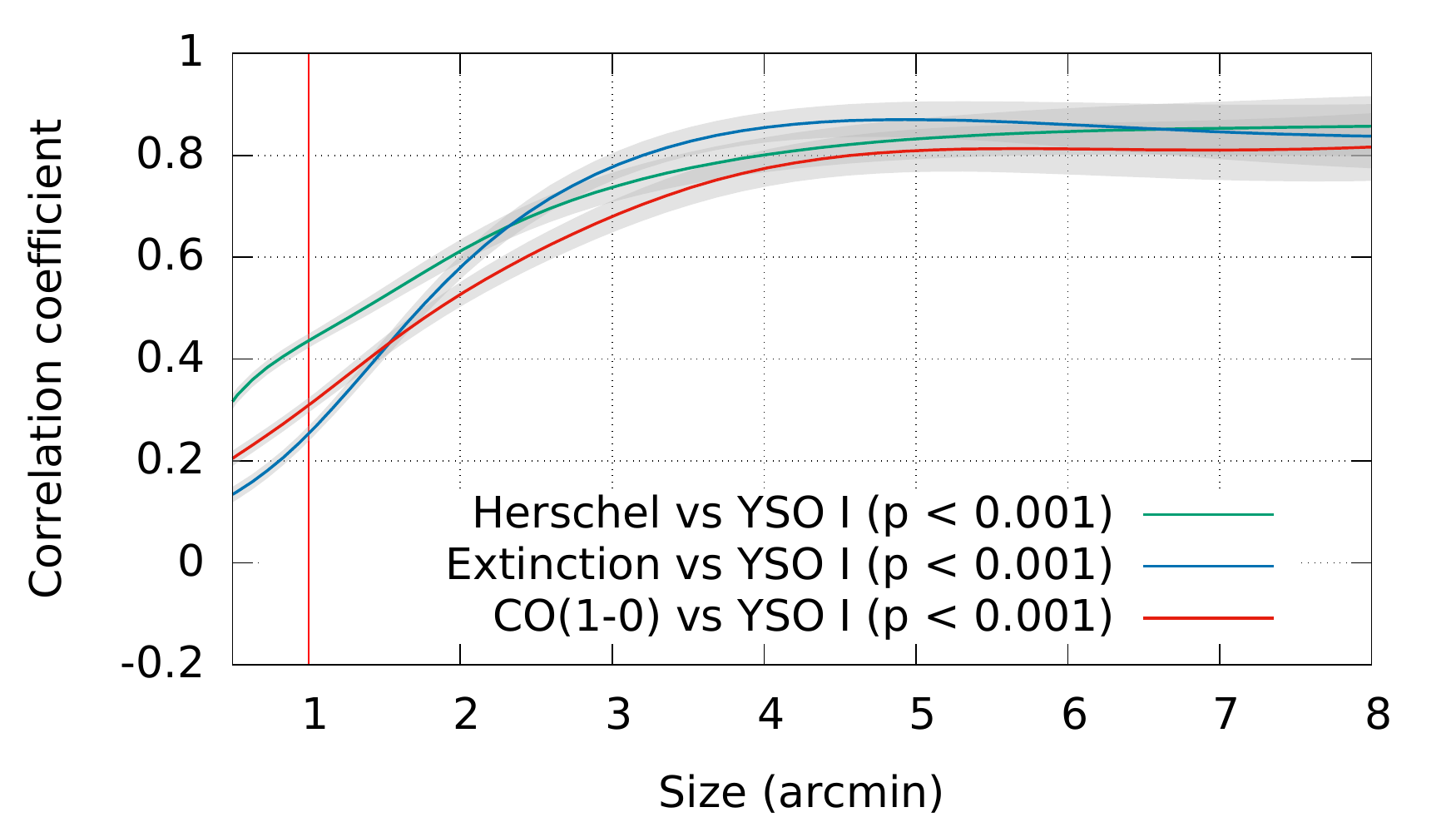}}\subcaptionbox{Class II YSOs (Whole region)}{\includegraphics[width=0.47\textwidth]{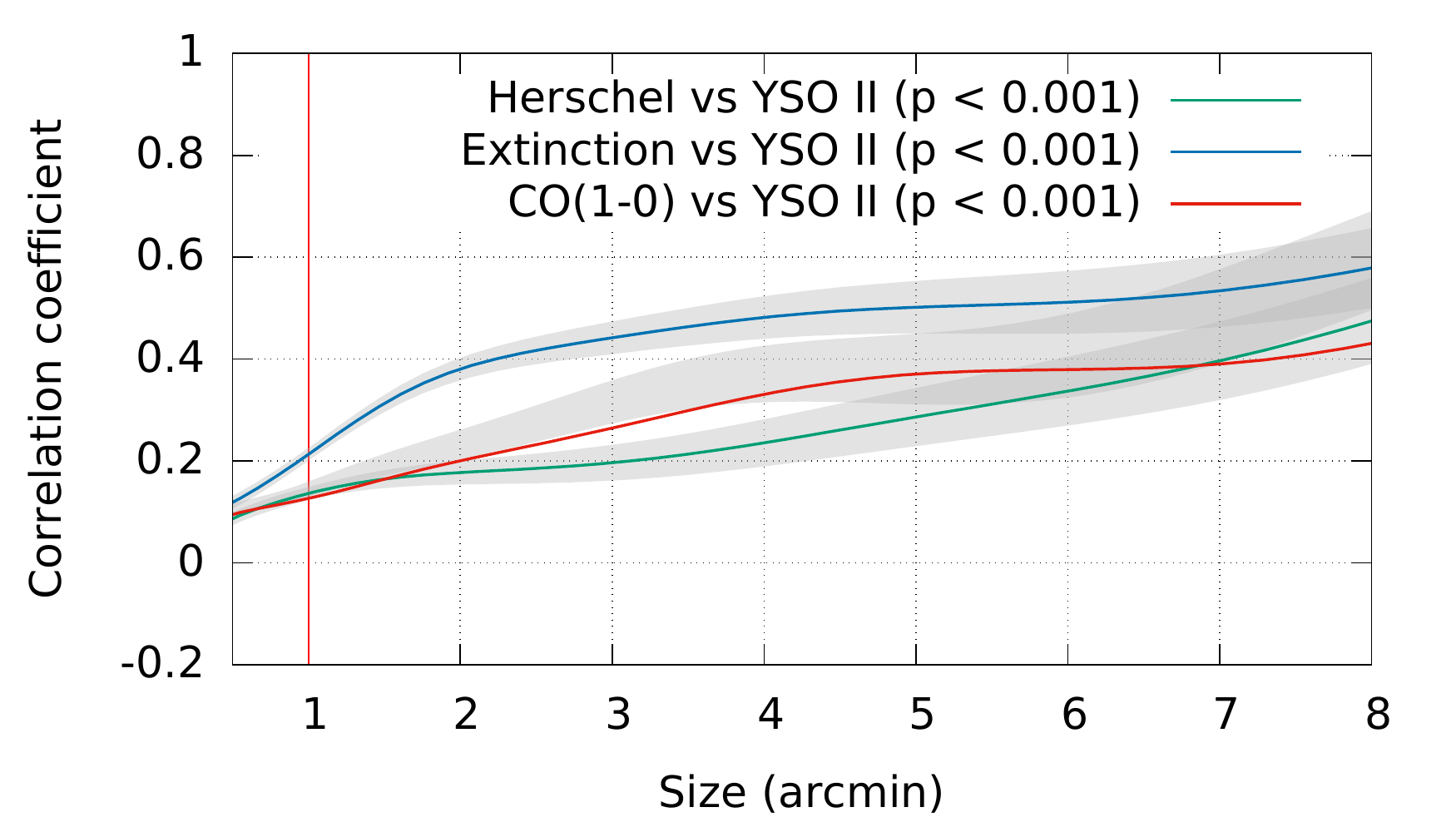}} \vspace{3mm}

\subcaptionbox{Class I YSOs (Without S255N)}{\includegraphics[width=0.47\textwidth]{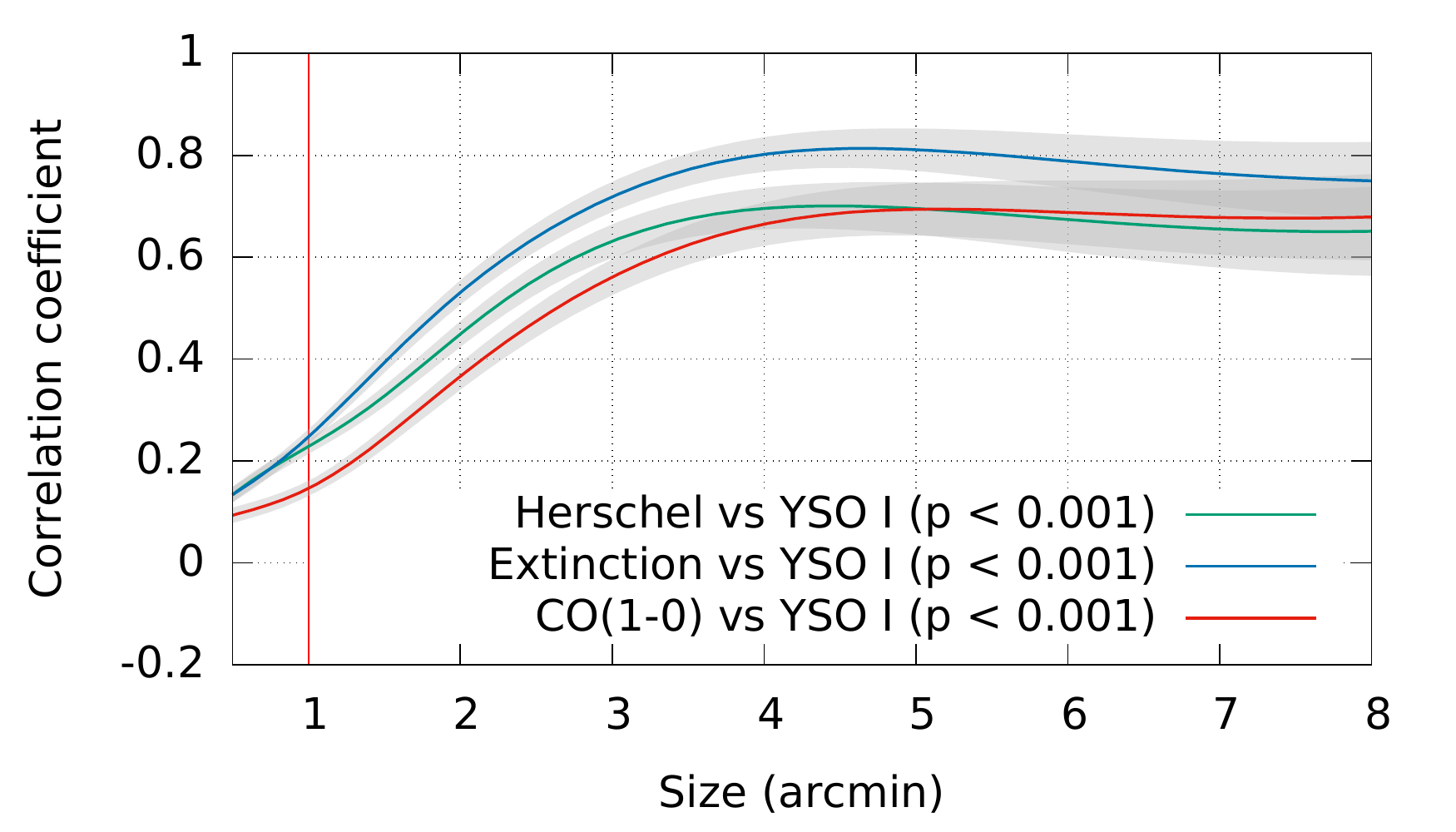}}\subcaptionbox{Class II YSOs (Without S255N)}{\includegraphics[width=0.47\textwidth]{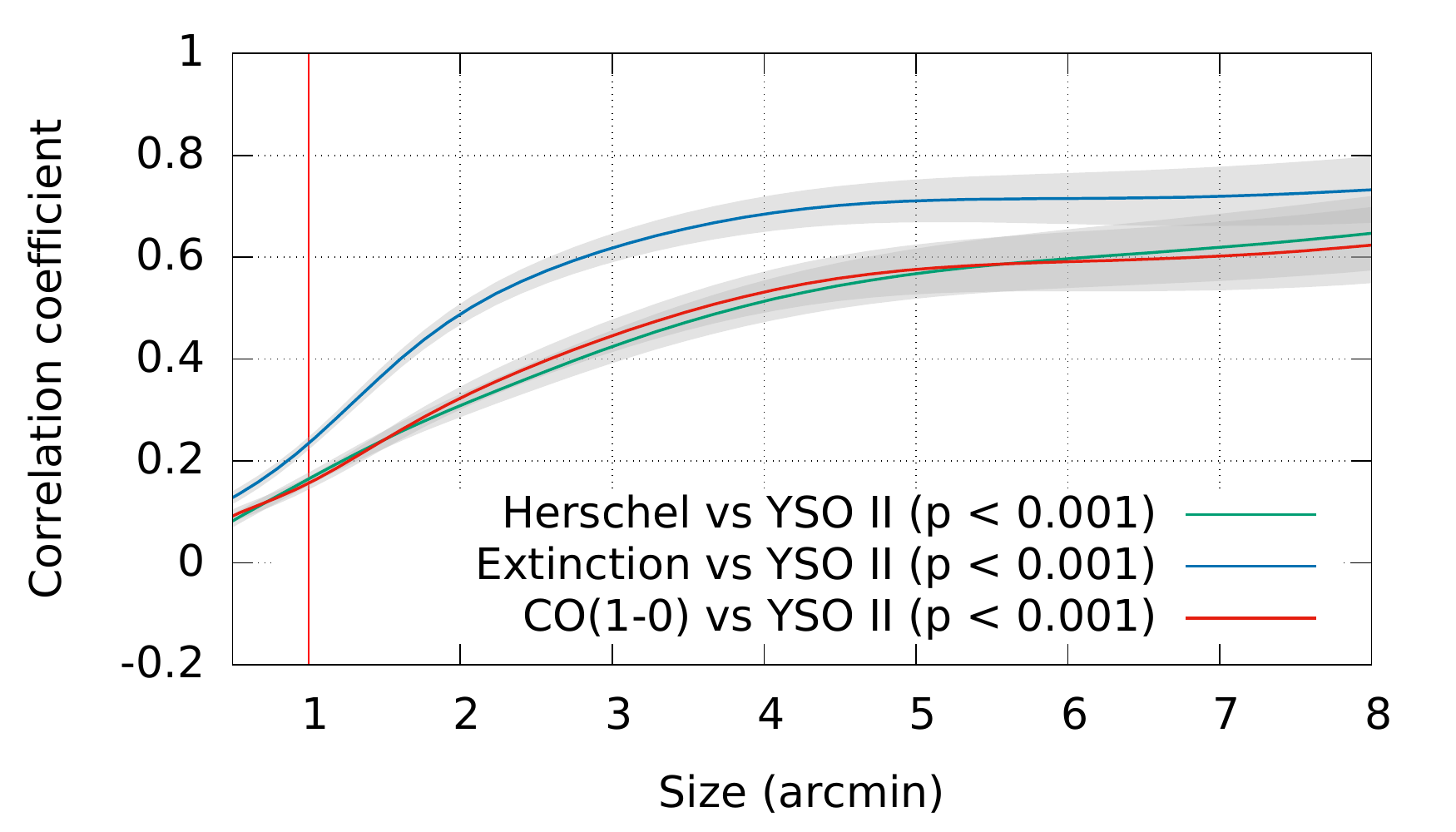}}\vspace{3mm}

\subcaptionbox{Class I YSOs (HCO+ mask)}{\includegraphics[width=0.47\textwidth]{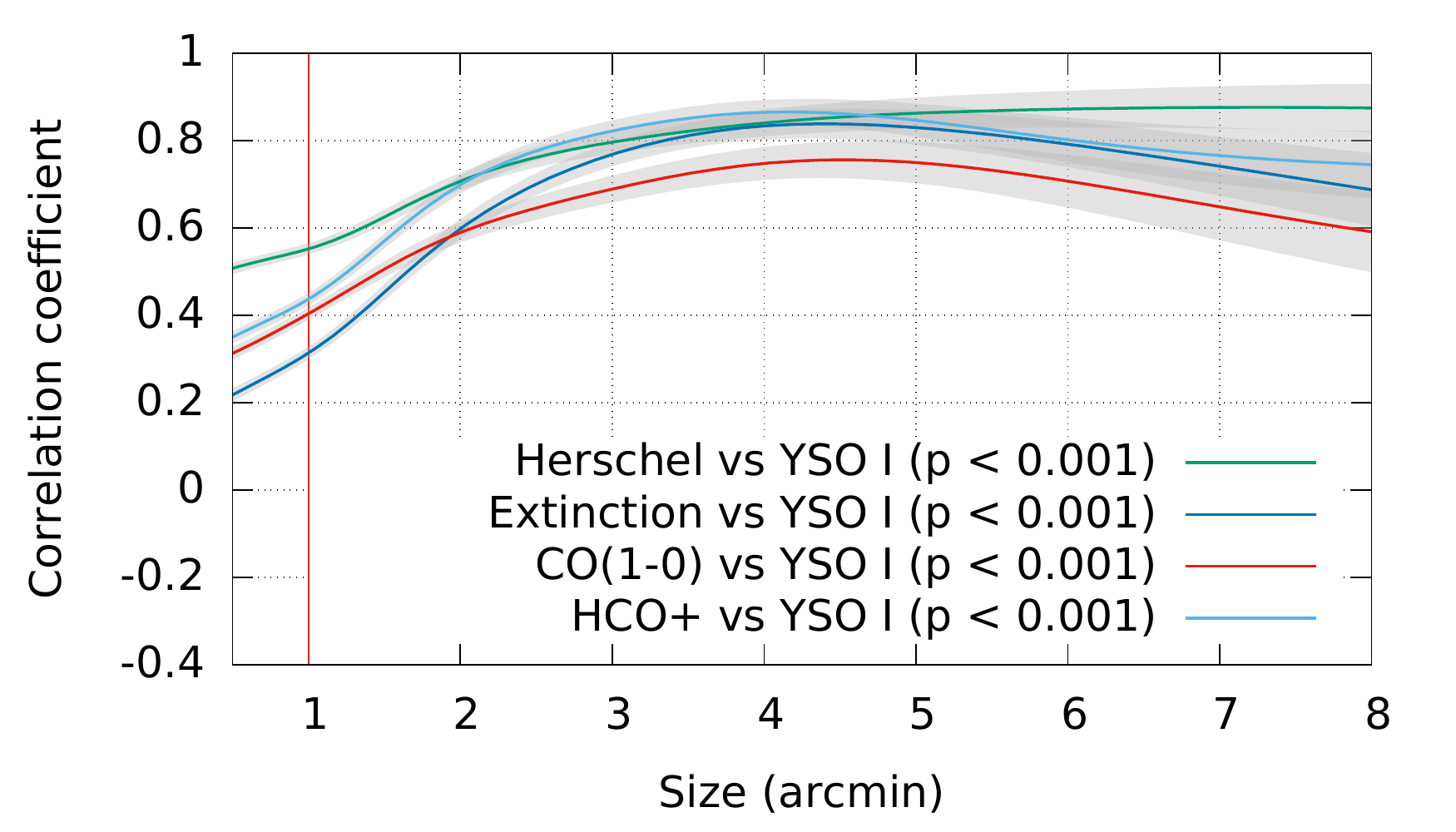}}\subcaptionbox{Class II YSOs (HCO+ mask)}{\includegraphics[width=0.47\textwidth]{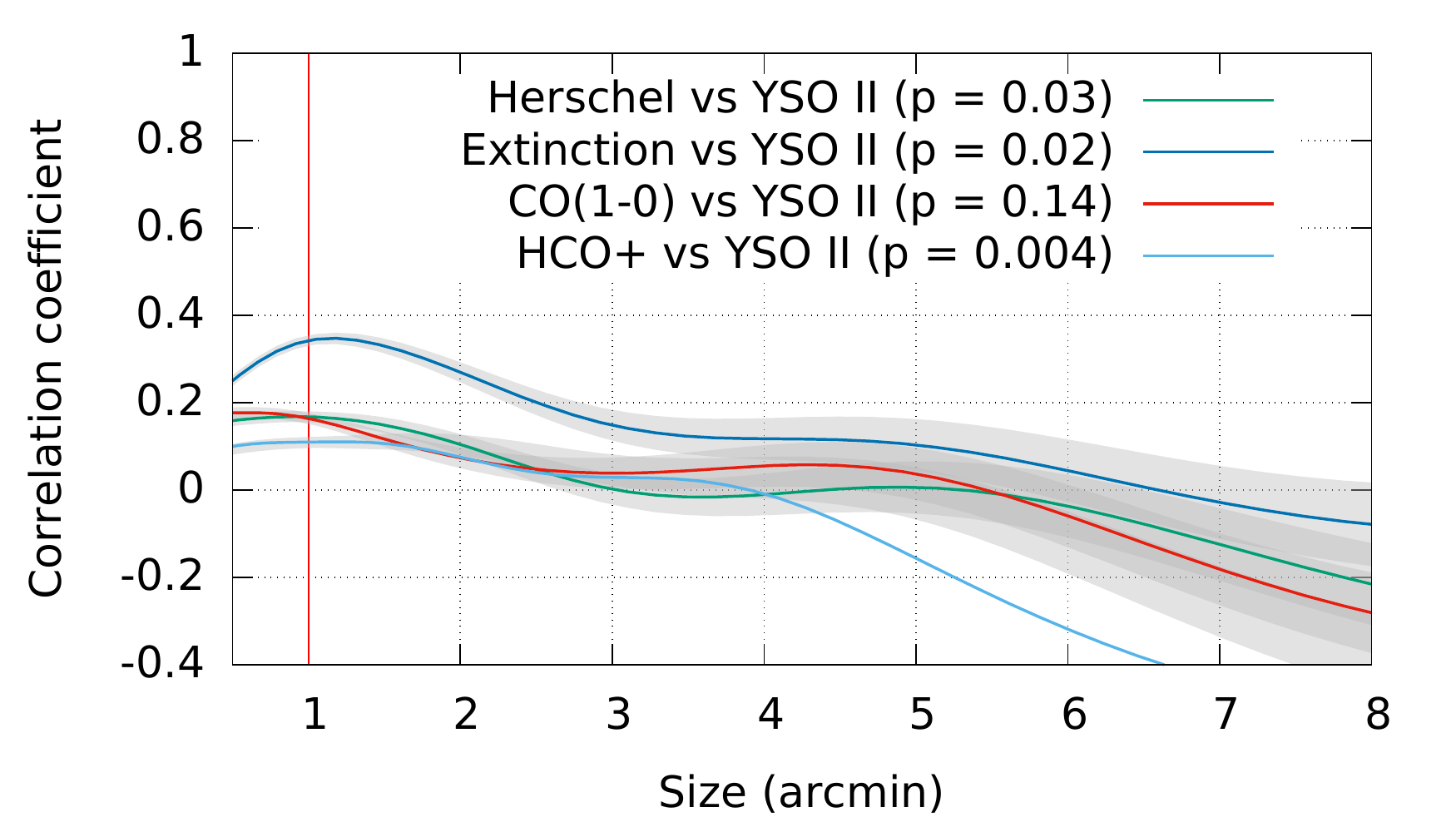}}
 \caption {Result of spatial correlation analysis between YSOs density and gas and dust tracers for the whole region (panel A,B), for the whole region without S255N cluster (panel C,D), and the \hco{} map masked region (panel E,F). The panels A, C, D display correlation coefficients with the class I YSOs, panels B, D, F display correlation coefficients with class II YSOs. The vertical red line in each plot displays all studied maps' beam size (60 arcsec). The gray-filled curve outline the error bars. Values of $p$ are results of the two-dimensional Kolmogorov-Smirnov (KS) test between WWCC analysis of the actual YSOs distribution and $N$ randomly distributed sources. One chose the lowest value of $p$ for all KS-tests with the different number of random sources ($N=50,100,150,250,500$).} 
 \label{fig:wwcc_correlation_results}
\end{minipage}
\vspace{5mm}
\end{figure*}

To compare the different scales of YSOs density with gas and dust tracers, we used the YSO density map with a gaussian kernel width of 60 arcsec (see details in the Section~\ref{YSO_map}). In order to distinguish between actual correlations and the artifact of comparing the mapped structures with the YSO density maps with increasingly large kernels we run the ``null'' test. We create the random distribution of YSOs in the studied region with the same number of sources as actual YSO catalog and create the 60 arcsec density map to test the behaviour of spatial correlation analysis.  One can then compare the test density map with the extinction, CO and Herschel maps.  The results of the spatial correlation analysis together with the ``null'' test using the WWCC method are presented in the Figure~\ref{fig:wwcc_correlation_results}. 

\subsubsection{Class I YSOs}

In the panel A of the Figure~\ref{fig:wwcc_correlation_results} we present the correlation between class I YSOs density and different gas and dust tracers for the whole studied region. At small scales (up to $\sim$2 arcmin) CO and Herschel maps tracers reveal  similar values of $r=0.5-0.6$. The extinction map at small scales have smaller values of $r$. The lower correlation between YSO density and extinction at small scales may be related to the larger relative noise of the extinction measurements. As we convolve other maps to the 60 arcsec resolution of the extinction map, they have a much smoother appearance. The uncertainty of the extinction measurements also increases towards high column densities, as the higher cloud opacity decreases the density of observed stars. This could also bias the extinction measurements, as stars are preferentially detected in those parts of the averaging beam where the column density values are the lowest \citep{Juvela16}. The other reason may be the shift of the extinction peak in the S255N region to the south compared to a similar peak in the CO and Herschel maps. The class I YSOs density map has a peak position similar to CO and Herschel maps, but the extinction map reveals the peak shifted to the south. For larger scales (> 2.0 arcmin) all maps have almost similar values of $r$ within the error bars. 

In the panels C and E, the correlation analysis within the masked regions is presented. In the ``no S255N region'' (panel C) the values of $r$ for extinction map is increased in comparison to CO and Herschel maps. Thus for the gas surrounding the S255N region the extinction map better represent the distribution of class I YSOs in comparsion to CO and Herschel maps. The direct pixel-by-pixel correlation analysis from the Section~\ref{sec:correlation_analysis} reveal a similar conclusion -- the extinction map in the ``no S255N region'' have the largest value of $r=0.52$ among other tracers. From the correlation analysis we found that this difference related to scales from 1 to ~4 arcmin.

From the analysis of Figure~\ref{fig:wwcc_correlation_results} (panels A,C) we conclude that for class I YSOs there is the `saturation point' of $\sim2-2.5$ arcmin. After reaching of this scale the correlation coefficients between column density and class I YSOs density do not increase significantly. We discuss the implication of this scale in the Section~\ref{history}. The scale of $\sim2-2.5$ arcmin is confirmed in all three studied cases: full region (panel A), excluding the S255N region (panel C) and the ``\hco{} mask'' region (panel E). In contrast, for class II YSOs the distribution of correlation coefficients is much smoother and contains no saturation point (see panels B,D of Figure~\ref{fig:wwcc_correlation_results}). Thus it is not obvious at which spatial scale the correlation becomes more pronounced.

Comparing the correlation coefficients between class I and II YSOs density and gas-dust tracers (panels A-D) we note that the extinction map has higher correlation coefficients with YSOs density of both class I and II at large scales (> 1.5-2 arcmin) in comparison to all other tracers. That is especially pronounced in the panels B, C and D of the Figure~\ref{fig:wwcc_correlation_results}. Comparing the YSOs class I and II density and extinction map visually (see Figure~\ref{fig:yso}), we found that all regions that have peaks of class I and II YSOs density (S255N, S256-south, G192.75-0.08, S258) similarly have the peak of extinction. Thus we conclude that extinction maps is a good tracer of both less evolved class I YSOs and more evolved class II YSOs.

The panels E and F of the Figure~\ref{fig:wwcc_correlation_results} present the result of the spatial correlation analysis under the  ``\hco{} mask'' region. From the analysis of the panel E we conclude that \hcoi{} map have values of $r$ with class I YSOs density similar to other tracers of gas and dust within the statistical error. This may give an idea of distribution of \hcoi{} line emission in the whole unmasked region -- it should generally follow other gas and dust tracers, especially at large scales (> 1.5 arcmin). We also note that at very large scales (4-8 arcmin) values of $r$ are decreased from 0.8 to 0.7.  This decrease may be associated with the influence of class I YSOs that are outside of the \hco{} mask coverage. Much stronger decrease is pronounced between class II YSOs and various gas and dust tracers (see panel f of Figure~\ref{fig:wwcc_correlation_results}). The number of class II YSOs that fall outside of \hco{} mask coverage region is very large, thus we may see even negative values of $r$. 

\subsubsection{Class II YSOs}

The correlation analysis between Class II YSOs and gas-dust tracers is presented in Panels B and D of Figure~\ref{fig:yso}. The correlation of Class II YSOs is much less pronounced with all studied gas and dust tracers when comparing to the Class I YSOs (panels  A, C). Even from  visual inspection of Figure~\ref{fig:yso}, one can see that the more evolved Class II YSOs are located at some distance from the regions of high gas column density in S255N and S256-south regions. 

If we consider the full studied region (panel B), the best correlation with class II YSOs is found in extinction map. The situation is not changed significantly if we exclude the S255N region and study the gas surrounding S255N (panel d) -- extinction maps have the highest $r$ compared to other maps. However, the absolute values of $r$ for class II YSOs are lower than for class I YSOs.  We note that for the CO map, the values of $r>0.5$ for class II YSOs become pronounced only at quite large scales (> 4 arcmins) and only when considering the mask excluding the S255N region (panel D). For the extinction map, the value of $r>0.5$ reached at the smaller scale of 2 arcmins when considering the mask excluding the S255N region (panel D). As the S255N region contain a small number of class II YSOs, but column density is significant, excluding that region results in much higher values of $r$ for all tracers. For class I YSOs values of $r=0.5-0.6$ are pronounced at much smaller scales (see panel C) starting from 1.5 arcmins in the unmasked region. We conclude that class II YSOs are located at a significant distance from the high column density regions and have only a weak relation to the maps considered here. 

Correlation analysis of various gas tracers with class II YSOs in \hco{} mask (panel F of the Figure~\ref{fig:wwcc_correlation_results}) reveal very low and even negative values of $r$. This may occur because significant part of class II YSOs reside in the regions outside the \hco{} mask region. Thus we conclude that it is difficult to study the spatial correlation between class II YSOs and gas and dust tracers in the \hcoi{} masked region. However, direct pixel-by-pixel analysis may be used in this case, as it has no influence of class II YSOs that are outside of the mask. 

From panels B and D, it is notable that increasing the spatial scale between class II YSOs and gas-dust tracers results in continuously increasing correlation coefficients. We associate this increase with the class II YSOs' proper motions -- during their evolution, the class II YSOs have enough time to fly away from the high-density region. Thus we can see the increase in correlation coefficients when increasing the spatial scale. However, increased correlation coefficients at large scales may be associated with map similarity due to convolution to large kernel size. In the next section, we will use the test maps to check the difference between actual YSOs and random sources distribution while comparing them with gas-dust tracer using the WWCC method.

\subsubsection{Comparsion with test maps}

To check the plausibility of comparing actual YSOs density and gas-dust tracers, we run several tests with the randomly distributed sources. These maps were analyzed using the WWCC method together with the gas-dust tracers (see Appendix~\ref{app_test_wwcc}). The two-dimensional Kolmogorov-Smirnov (KS) test was used to check the difference between the actual YSOs correlation with gas-dust tracers and correlations with the $N$ random sources density map.  The different values of $N$ have been investigated: 50, 100, 150, 250, and 500 sources. 

The resulting $p$-values are presented in the legend of Figure~\ref{fig:wwcc_correlation_results} and results of WWCC analysis using the random sources maps presented in Figure~\ref{fig:random_YSOs} of Appendix~\ref{app_test_wwcc}. The KS test reveals a statistically significant difference between correlation coefficients of gas-dust tracers with class I/II YSOs and randomly distributed sources. While using the extinction map for WWCC analysis, the difference between class I and class II YSOs and random sources distribution is statistically significant in full and ``no S255N'' regions, i.e., the null hypothesis about the same distribution can be rejected at a 3$\sigma$ level (p < 0.001). In the ``HCO$^+$ mask'' region, the difference between Class I YSOs and random is also significant at $3\sigma$ level (p < 0.001). The exceptional case is class II YSOs and map with $N=$150 sources in the ``HCO$^+$'' mask region -- the KS test gives a $p$-value of 0.0018. Thus one cannot reject a null hypothesis about the same distribution at a 3$\sigma$ level ($p$ > 0.001). 

The same behavior is found for the CO and Herschel maps: both class I and class II YSOs density maps are statistically different from random sources distribution except for the ``HCO$^+$'' mask region, where class II YSOs density map correlations is similar to the random map of 100 sources ($p$ = 0.14). We conclude that both class I and class II YSOs display a statistically significant correlation with gas-dust tracers, not related to the effect of comparing large-scale structures. However, the ``HCO$^+$'' mask region has insufficient spatial coverage to display statistically significant differences between actual YSOs position and random sources.

\begin{figure*} 
\vspace{5mm}
 \center
\includegraphics[width=0.99\textwidth]{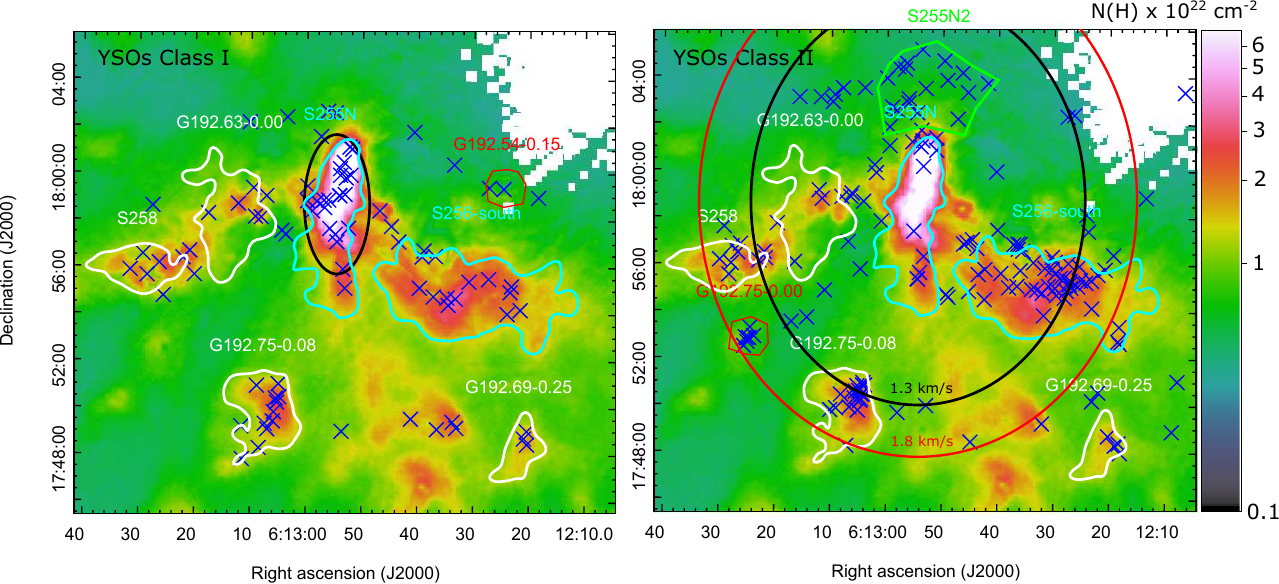}
 \caption {Distribution of class I (left panel) and class II (right panel) YSOs in the S254-S258 star formation region. YSOs are marked with blue crosses. The black ellipse in the left panel represents the relative span of class I YSOs in the S255N cluster, while the red ellipse on the right panel represents the span of class II YSOs using the expansion velocity of 1.8 km$\,$s$^{-1}$ found from the dispersion of HCO$^{+}$ line. The black ellipse in the right panel displays class II YSOs' span using an expansion velocity of 1.3 km/s found in \citet{Bieging16} for the S235-AB region. }
 \label{fig:yso_herschel}
\vspace{5mm}
\end{figure*}

\section{History of the star formation in the S255 region} \label{history}

From the direct comparison between gas and dust column densities and YSOs density (see Section~\ref{correlation_power_law}), it was shown that the density of class I YSOs has a positive power law relation to the column density with an average slope of $\sim1.2$ and a correlation coefficient of $\sim0.6-0.7$ for all studied gas and dust tracers. In comparison, the slope for class II YSOs is $\sim0.65$ and the correlation coefficient is only $\sim0.3-0.4$. This difference may be related to the strong association between young stars and the primordial gas where these stars were born, traced by the different gas and dust tracers. 

More evolved class II YSOs display less pronounced power law relations and lower values of correlation coefficients due to the greater age of class II stars, as they have managed to move away from the primordial gas. At small scales (< 4 arcmin) the correlation between class II YSOs and column-density tracers is poor ($r=0.2-0.4$) because the stars have already left their natal clumps, but at larger scales (> 4 arcmin) some correlation remains ($r=0.5-0.6$) because the YSOs have not yet moved far enough. 

From the spatial correlation analysis, we found that at $\sim2-2.5$ arcmin, there is the `saturation point' where correlation coefficients between class I YSOs and various gas and dust tracers began to grow much more slowly. At the assumed distance of 1.59 kpc, this spatial scale corresponds to $\sim$0.9-1.1 pc. From the typical duration of class I YSOs stage (0.5 Myr according to \citet{Heiderman2015}), the average YSOs travel speed is then $\sim$1.8-2.3~km~s$^{-1}$.  The work of \citet{Bieging16} argue that the spatial distribution of YSOs of Class 0/I compared with Class II is consistent with the young stars having the same velocity dispersion as that which we observe in the gas for the highest column density regions. From the $^{13}$CO(2-1) line in the direction of the S255N region, we found the median line dispersion of 1.4 km~s$^{-1}$ with the maximum value of 1.86 km~s$^{-1}$. A similar analysis for \hcoi{} line give median dispersion of 1.3 km~s$^{-1}$ and maximum dispersion of 1.7 km~s$^{-1}$. Thus the travel speed of $\sim$1.8 km~s$^{-1}$ is consistent with the $^{13}$CO(2-1) velocity dispersion in the direction of high column density regions. However, the actual YSO movements may be slower than the gas dispersion and the reaching of the spatial scale turning point. Thus we assume 1.8 km~s$^{-1}$ as an upper limit.

We analyzed YSOs evolutionary stages duration in S255 region similar to \citet{Bieging16}. We enclose the Class I YSOs in the S255N region in the ellipse with the major and minor axes of 2.97$\times$1.39\arcmin{}. At the distance of 1.59 kpc, this corresponds to 1.38$\times$0.64~pc. Using a travel speed of 1.8 km~s$^{-1}$ and duration of class II YSO stage of 2 Myr \citep{Heiderman2015} we conclude that due to random YSO motions, the ellipse of class II YSOs should be increased by 3.67 pc in both directions that lead to dimensions of 4.32$\times$5.05~pc. Figure~\ref{fig:yso_herschel} reveal the initial ellipse on the left panel and expanded ellipse using the expansion velocity of 1.8 km~s$^{-1}$ in the right panel. We additionally put the ellipse for class II YSO using the expansion speed of 1.3 km~s$^{-1}$, similar to the one used in \citet{Bieging16}. From the analysis, we found that the almost all YSOs clusters observed in the S254-S258 region are enclosed in the expanded ellipse of 1.8 km~s$^{-1}$. The situation is not changed significantly if one used the speed of 1.3 km~s$^{-1}$, as found by \citet{Bieging16} for S235-AB region.

In the classical picture of star formation, the gas is  concentrated into a single region, stars forming in situ, and the resulting cluster then dispersing.  In the S254-S258 star-formation region, the absolute peak of class I YSOs density was found toward the S255N cluster, which also has the absolute maximum of gas column density. Thus we can assume that this cluster may be the starting point for many YSOs that already have left its birthplace. Starting from the S255N, YSOs have enough time to move through the whole S254-S258 region. Given that the whole star formation region contains several isolated YSOs clusters, the evolutionary link between the S255N cluster and the other YSOs clusters can be considered. A possible example of YSOs movement is seen in the S255N and S255N2 clusters. As we observe many class I YSOs and lack of class II YSOs in the S255N region, we conclude that YSOs are quickly flying away from the S255N region. However, YSOs are dispersed not in an isotropic way -- most of them have moved toward the S255N2 region.

In the S255N2 region many of class II YSOs are observed, but gas and dust emission was not detected in all considered tracers, including CO, $^{13}$CO, CS, HCO$^+$, near-IR extinction and Herschel maps.  The class II YSOs is much more scattered in S255N2 region in comparison to other regions rich in class II YSOs (S256-south, S258, G192.75-0.08, G192.75-0.0). Given that these regions harbour gas with high column density we suggest that interaction between gas and stars can create better conditions for stability of star clusters  in comparison to cluster that is just moved away from it's birth material (S255N2) -- clusters formed in deeper potential wells should disperse more slowly. However, the stability of YSOs cluster is not only about the gas mass but also about the mass in the cluster itself. The differences in the cluster structure could also be related to different dynamics during their formation.

However, with the available data, we can't resolve why Class~II YSOs are moving primarily to the north of the S255IR region and how other YSOs clusters (S256-south, G192.75-0.08, G192.75-0.00, S258, G192.63-0.00) are formed. It is not possible to determine whether star formation in the gas clumps is going continuously or episodically. Ultimately, Class II YSOs formed earlier in the same gas condensation and moved away while class I sources are formed more recently and stay close to positions of their formation. Moreover, the duration of class II YSOs stage ($\sim$2 Myr), and the gas velocity dispersion of 1.8 km~s$^{-1}$ allow one to travel through the whole S254-S258 region.

The \hco{} map reveals the regions of primordial gas where stars are being born. After young stars evolve they are moved to the class I and we can see relatively high values of $r=0.6-0.8$ between class I YSOs density and gas column density traced by extinction, CO and \hco{} maps at scales 1.5-3 arcmin. When YSOs enter the class II stage, they have already moved away from from the primordial gas, and we can find a good correlation ($r=0.4-0.6$) between class II YSOs and gas column density only at large scales (>4 arcmin), while small scales (< 4 arcmin) have low values of $r=0.1-0.4$. Thus the CO and extinction maps at large scales resemble the "old" star formation history seen in class II objects. In contrast, \hco{}, CO and extinction map at smaller scales (<1.5 arcmin) resembles the recent star formation history seen in class I YSOs. 

Concluding the view on the star formation in the S254-S258 region, the following picture comes out.  During the evolution the class II YSOs have enough time to move thought the whole S254-S258 region. Thus the clusters of Class~II YSOs in the S254-258 star-formation region can contain objects born in the different locations of the complex. The younger class I YSOs are found near it's primordial gas, thus we observe high correlations between class I and high column density regions. The more evolved class II YSOs are not directly related to gas column density at small scales. However, we found an continuously increasing correlation coefficient ($r=0.4-0.8$) between gas column density and class II YSOs density at large scales (4-8 arcmins). We associate this increase with the proper motions of class II YSOs -- during its evolution ($\sim$2 Myr) the class II YSOs have enough time to fly away to the maximum distance of 3.67 pc that corresponds to $\sim8$ arcmin at the distance of 1.59 kpc.

\section{Conclusions}

We have studied the star-forming region S254-S258, which includes several \hii{} regions and star clusters, in the different density tracers of molecular gas (CO, $^{13}$CO, HCO$^+$ and CS line emission) and dust (near-IR extinction and Herschel maps). The following conclusions are obtained:

\begin{enumerate}

\item High-density gas was detected in the inter-clump bridge between S255N and S256-south regions. Thus these regions may have an evolutionary link and trace different parts of the same molecular cloud.

\item Direct pixel-to-pixel correlation analysis revealed that less evolved class I YSOs are well-correlated with all studied gas and dust tracers with a correlation coefficient of $r=0.6-0.7$. On the contrary, class II YSOs have a larger spread and the lower correlation coefficients with the gas and dust tracers: $r=0.4-0.6$.

\item From the spatial correlation analysis it was found that the spatial size at which the correlations coefficients between gas column density and YSOs reach maximum values is $\sim$2 arcmins for class I YSOs. That corresponds to the scale of $\sim1.4$~pc and mean YSOs velocity of $\sim$1.8~km~s$^{-1}$ while assuming the distance of 1.59 kpc and duration of class I YSO stage of 0.5 Myr.

\item For class II YSOs, the spatial correlation coefficient increase continuously with the increase of the scale without a saturation point. We associate this increase with proper motions of class II YSOs.  According to the Kolmogorov-Smirnov test, this increase is statistically different from the correlation between gas-dust tracers and a random sources distribution.

\item In the direction of S255N region the velocity dispersion in the high column density regions is 1.8 km~s$^{-1}$ according to $^{13}$CO(2-1) line emission and 1.7 km~s$^{-1}$ according to \hco{} line emission. This velocity is consistent with the velocity found from the spatial correlation analysis.

\item Assuming the motion of YSOs with the speed of 1.8 km~s$^{-1}$, starting from the central S255N reigon, class II YSOs have enough time ($\sim$2 Myr) to reach the whole S254-S258 region, including isolated YSOs clusters. 

\item We found a non-isotropic distribution of class II YSOs around S255N that is shifted to the north and absence of class II YSO toward the S255N cluster. 

\item The spatial correlation analysis lead to the following conclusions: the \hco{} map reveals the regions of primordial gas where stars are being born. Class I YSOs have relatively high values of correlation coefficient $r=0.6-0.7$ with \hco{}, CO and extinction map at smaller scales (<2.0 arcmin), thus they resembles the recent star formation history. In contrast, the CO and extinction maps at large scales (> 4 arcmins) resemble the "old" star formation history seen in class II objects. 

\end{enumerate}

\section*{Acknowledgment}
 The work of M.~S.~K in the Section~\ref{hight_density_molecules} was supported by the RFBR grant 20-02-00643. The work by D.~A.~L. in Section~\ref{sec3} was supported by the Russian Ministry of Science and Higher Education, FEUZ-2020-0030. The work of S.~A.~K. of correlation analysis in Section~\ref{sec:correlation_analysis} was supported by the Russian Science Foundation grant 19-72-10012. The work of A.~M.~S in Section \ref{history} was supported by the Large Scientific Project of the Russian Ministry of Science and Higher Education ``Theoretical and experimental studies of the formation and evolution of extrasolar planetary systems and characteristics of exoplanets'' (No. 075-15-2020-780, contract 780-10).  V.~O. was supported by the Collaborative Research Centre 956, sub-project C1, funded by the Deutsche Forschungsgemeinschaft (DFG), project ID 184018867. The authors acknowledges support from Onsala Space Observatory for  the  provisioning of its facilities/observational support. The Onsala Space Observatory national research infrastructure is funded through Swedish Research Council grant No 2017-00648.

This work use data products from UKIRT Infrared Deep Sky Survey (UKIDSS) archive. The UKIDSS project is defined in \citet{Lawrence07}. UKIDSS uses the UKIRT Wide Field Camera \citep{Casali07}. The photometric system is described in \citet{Hewett06}, and the calibration is described in \citet{Hodgkin09}. The pipeline processing and science archive are described in \citet{Hambly08}.

\section*{Availability of data}

The data underlying this article are available in Figshare, at \href{https://dx.doi.org/10.6084/m9.figshare.c.5325083}{https://dx.doi.org/10.6084/m9.figshare.c.5325083} 



\bibliographystyle{mnras}
\bibliography{S255}



\appendix

\section{Conversion between the extinction and the gas column density} \label{app_extinction_h2}

The extinction maps described in the Section~\ref{sec:Extinction_maps} were used to estimate the column density in S254-S258 region. For the conversion between NIR (A$_J$) and visual (A$_V$) extinction we adopted the coefficients from \citet{Cardelli89}: $A_V=3.54\times A_{J}=5.26\times A_{H}=8.77\times A_{K}$. To  compute the hydrogen column density from the visual extinction we use the standard conversion of \citet{Bohlin78}:  $N(\mathrm{H})/E_\mathrm{B-V}=5.8\times10^{21}~(\mathrm{cm}^{-2}\mathrm{mag^{-1})}$. With the typical ratio of total to selective extinction $R_V=A_V/E_\mathrm{B-V} \simeq 3.1$ \citep{Cardelli89} this gives a hydrogen column density: $N(\mathrm{H})/A_V =1.87\pm0.13\times10^{21}~(\mathrm{cm}^{-2}\mathrm{mag^{-1})}$

However, in calculation of the gas column density from the extinction, the conversion factor between $N(H)$ and $A_V$ include some uncertainty. Different values of this ratio were obtained using far-UV extinction observations \citep{Diplas94}, X-ray sources observations  \citep{Predehl95,Guver09,Zhu17} and interstellar absorption-line observations \citep{Gudennavar12}. In the work of \citet{Gudennavar12} a survey of $\sim$3000 stars in the Galaxy were used to determine the conversion factor for total hydrogen (in molecular and atomic form) and extinction. The resulting value of $N(\mathrm{H})/A_V=1.96\pm0.06\times10^{21}$~cm$^{-2}$mag$^{-1}$ (assuming $R_V=3.1$) is close to the canonical value discussed above. The recent work of \citet{Zhu17} gives estimation of $N(\mathrm{H})/A_V=2.08\pm0.02\times10^{21}$~cm$^{-2}$mag$^{-1}$.

\section{Spatial correlation analysis tests} \label{app_test_wwcc}

In this section we examine the behavior of the WWCC analysis between gas-dust tracers and a random YSOs density map. This is nessesary to ensure that found correlation coefficients between actual YSOs and gas-dust tracers are not related to the properties of the WWCC method. In particular, we examine the increase of correlation between large-scale convolutions of the maps that actually have no relation to each other.

\begin{figure*}
\centering
\includegraphics [scale=0.45] {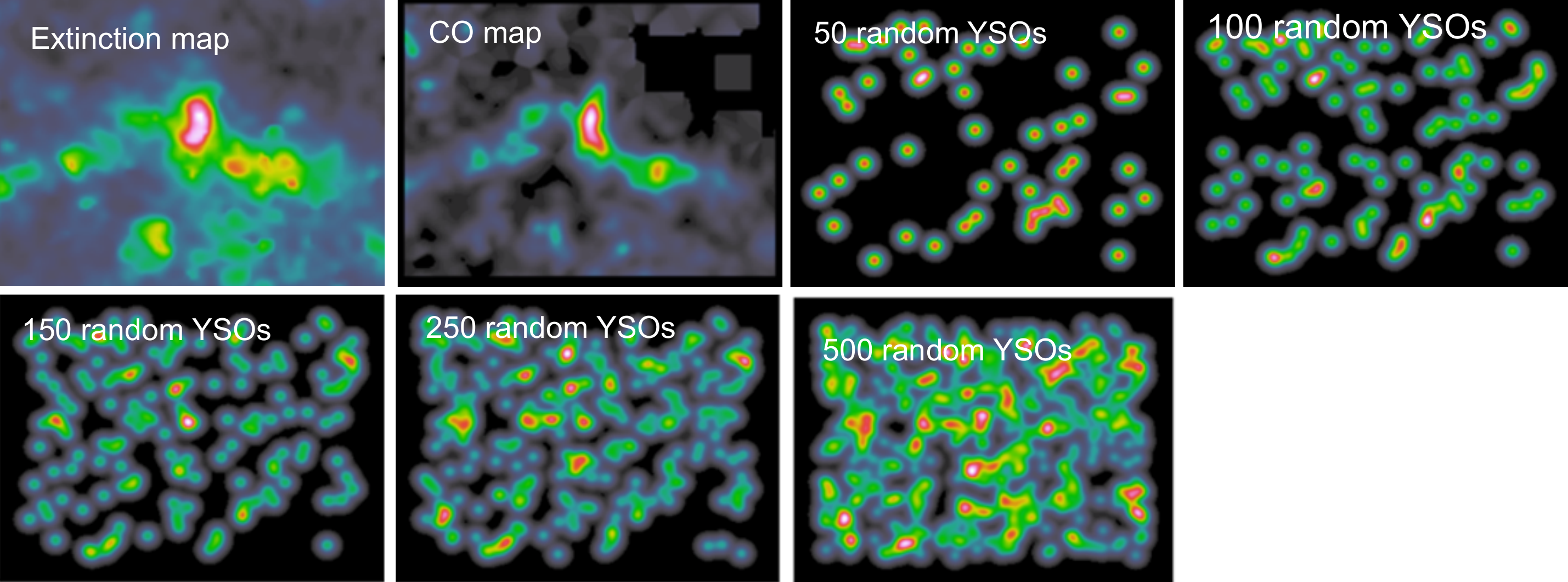}
\caption {Extinction, CO(1-0) and random sources density maps used for WWCC method testing.}
 \label{fig:random_YSOs}
\vspace{5mm}
\end{figure*}

The random YSOs maps are presented in Figure~\ref{fig:random_YSOs} together with CO and extinction maps that we used in the WWCC method testing. The following number of random YSOs were examined: 500, 250, 150, 100, and 50. We generated the catalog of random YSOs only once, and each time the limit on the source number was applied. Each YSOs density map was created using a kernel size of 60 arcsec, similar to the CO and extinction map used in the analysis.

\begin{figure*} 
\vspace{5mm}
 \begin{minipage}{180mm}
 \center
\subcaptionbox{Random sources test (Whole region)}{\includegraphics[width=0.4\textwidth]{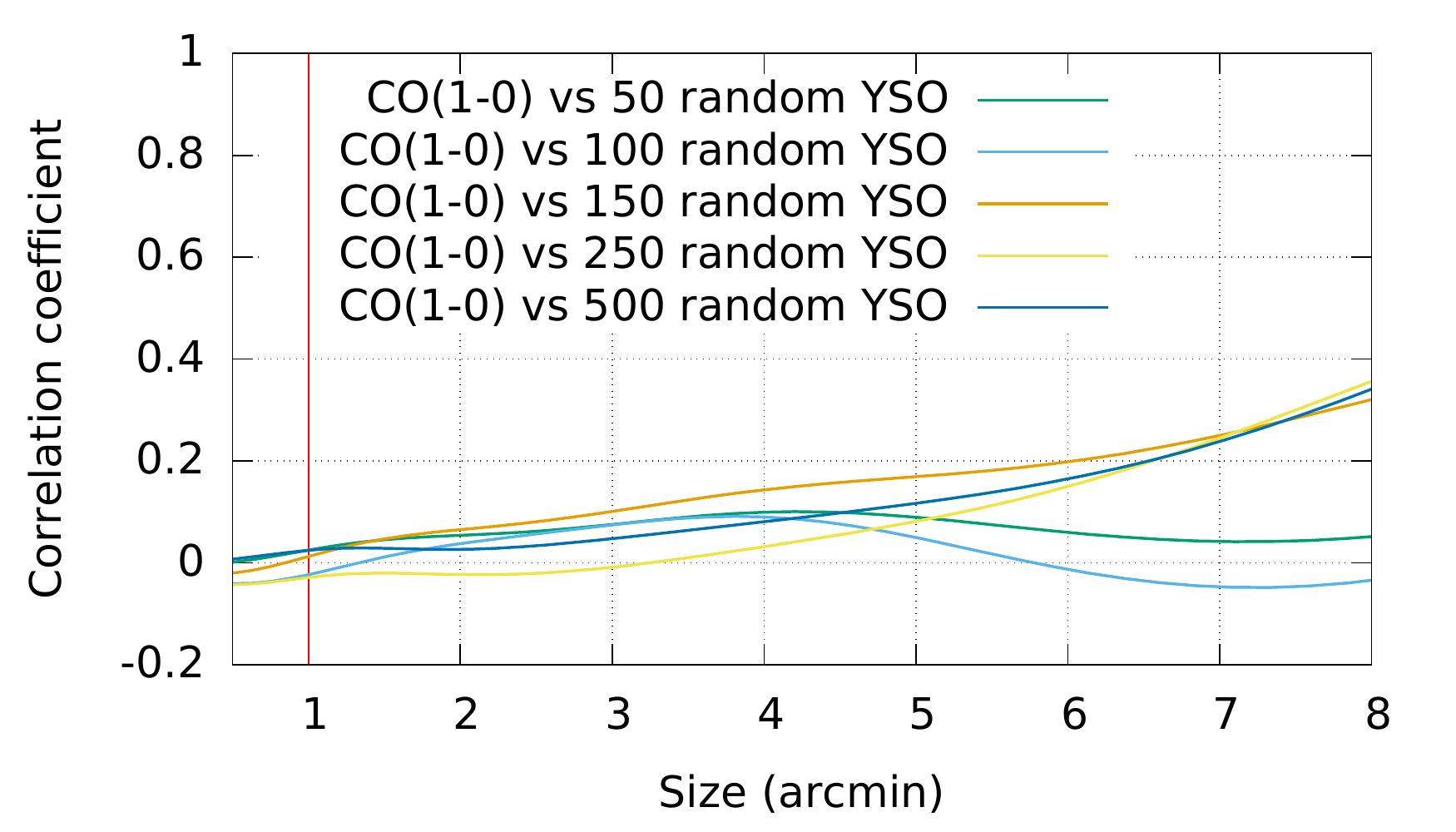}}\subcaptionbox{Random sources test (Whole region)}{\includegraphics[width=0.4\textwidth]{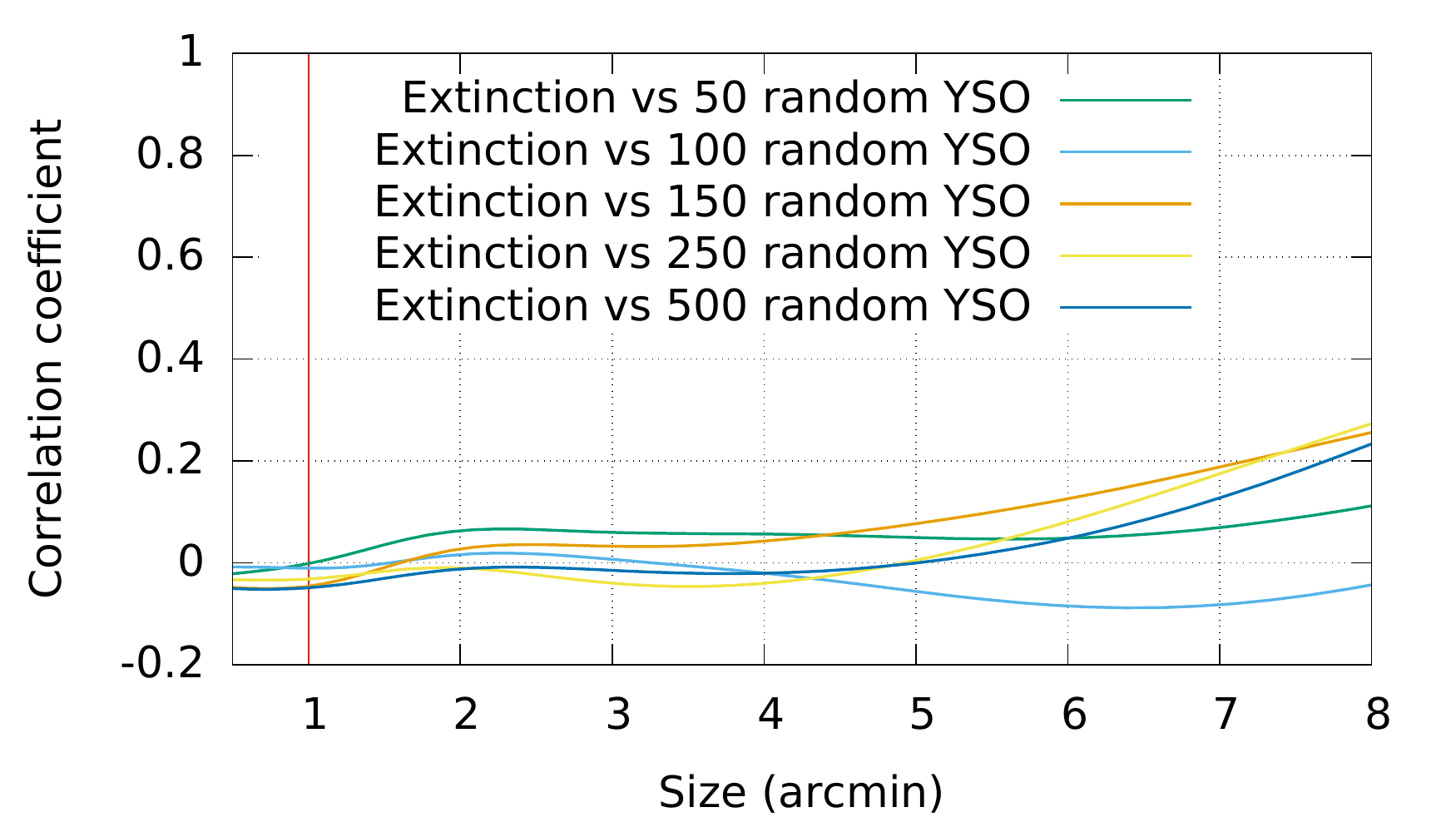}} \vspace{2mm}

\subcaptionbox{Random sources test (Without S255N)}{\includegraphics[width=0.4\textwidth]{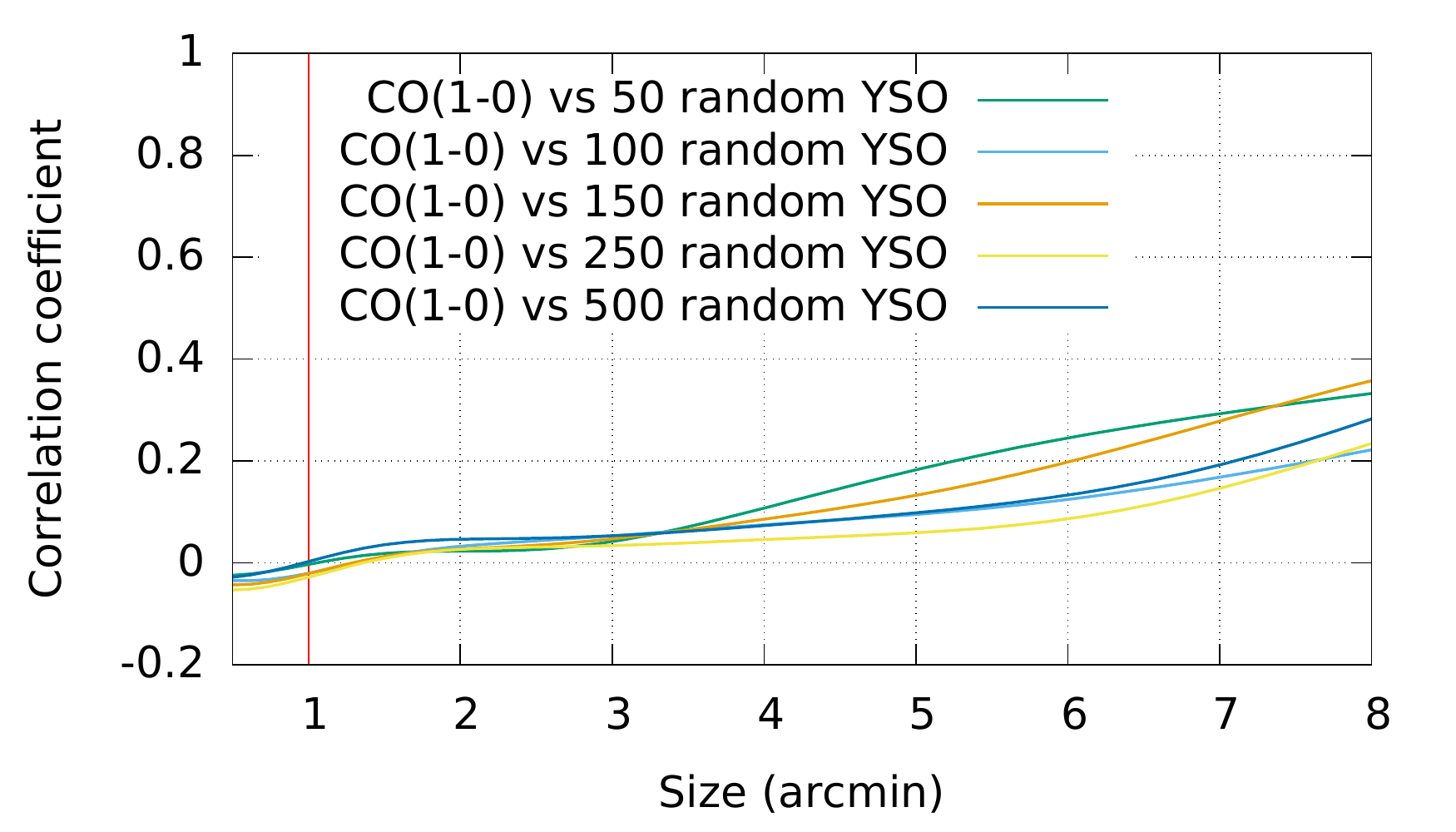}}\subcaptionbox{Random sources test (Without S255N)}{\includegraphics[width=0.4\textwidth]{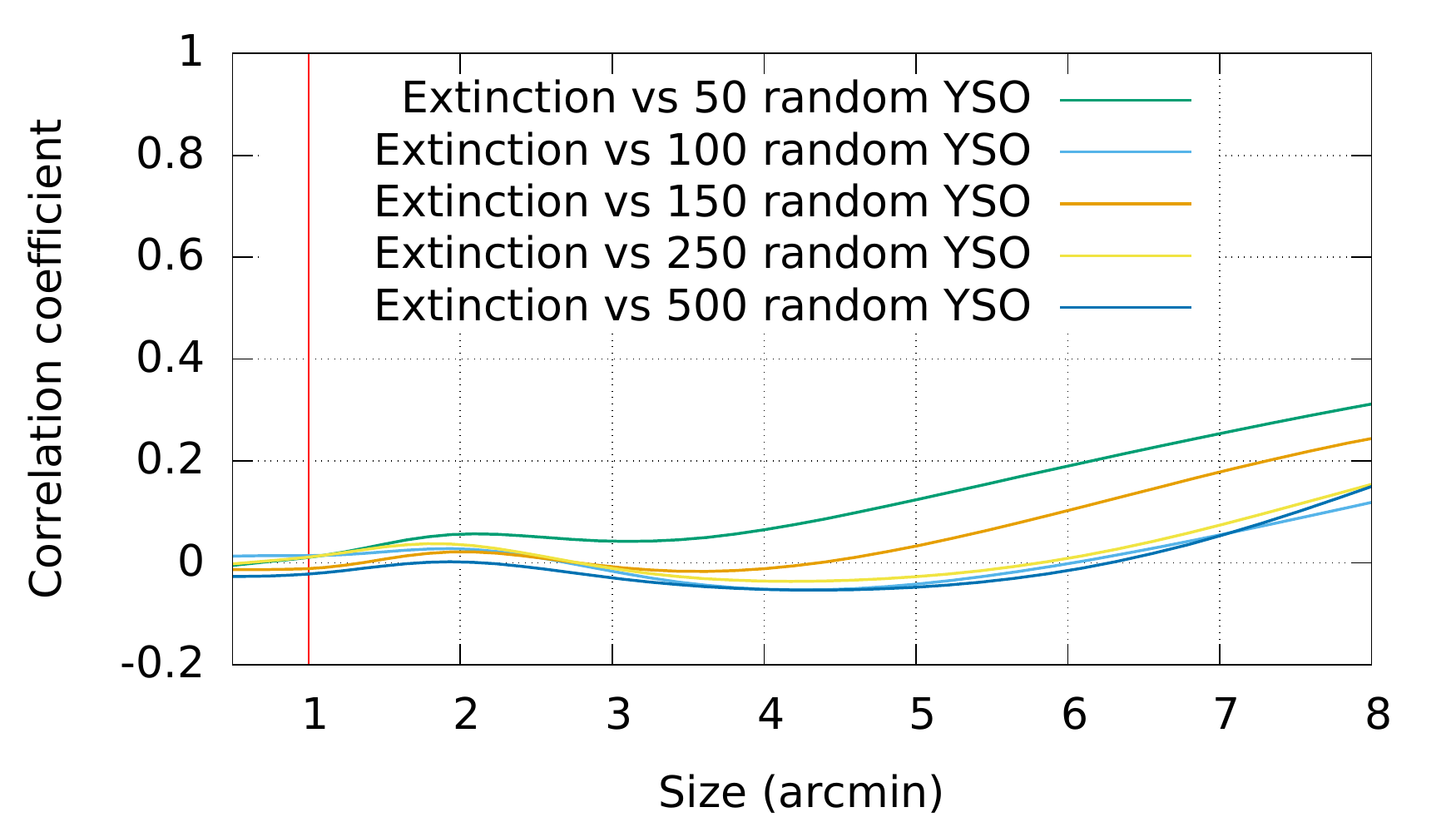}}\vspace{2mm}

\subcaptionbox{Random sources test (HCO+ mask)}{\includegraphics[width=0.4\textwidth]{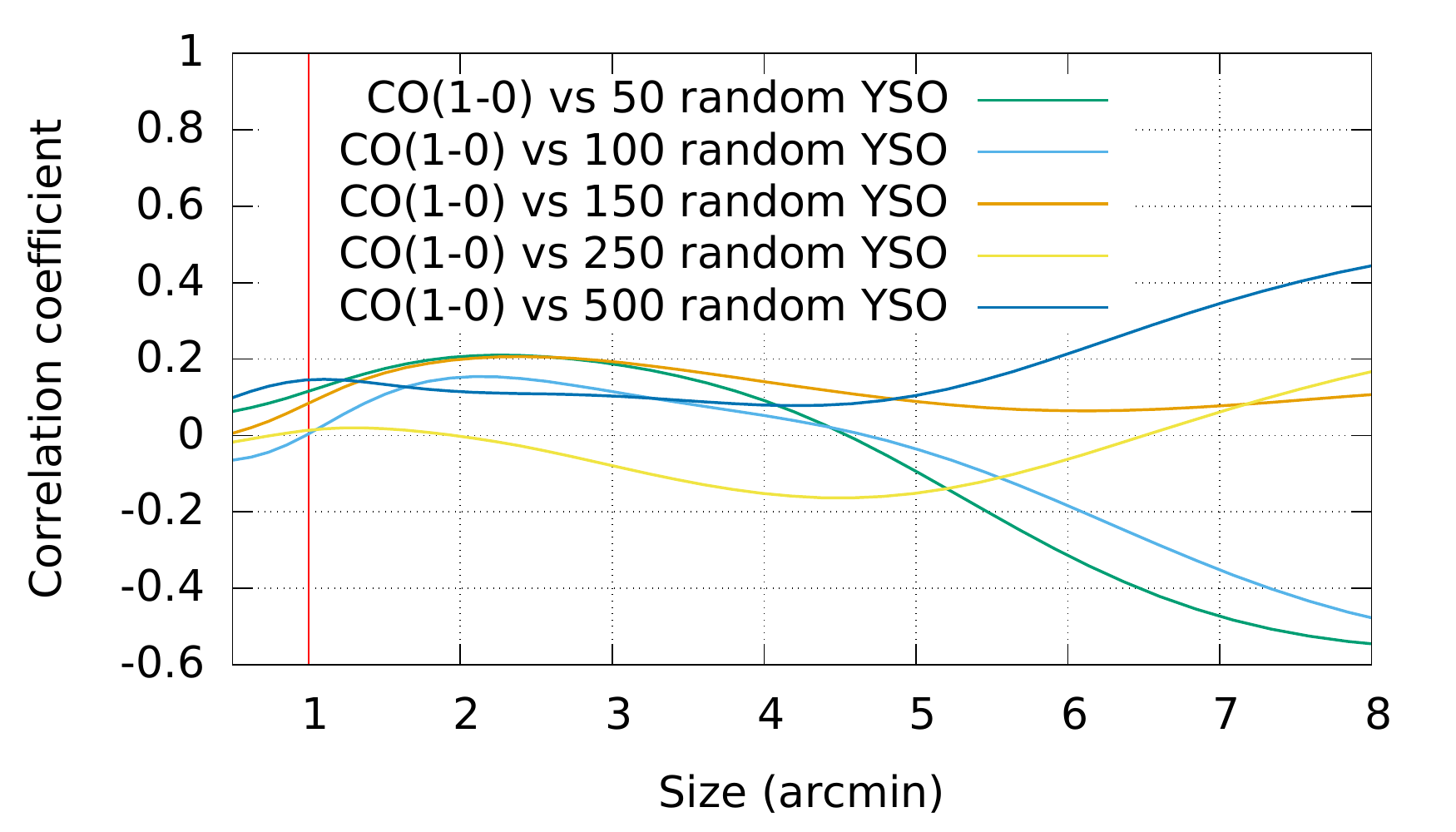}}\subcaptionbox{Random sources test (HCO+ mask)}{\includegraphics[width=0.4\textwidth]{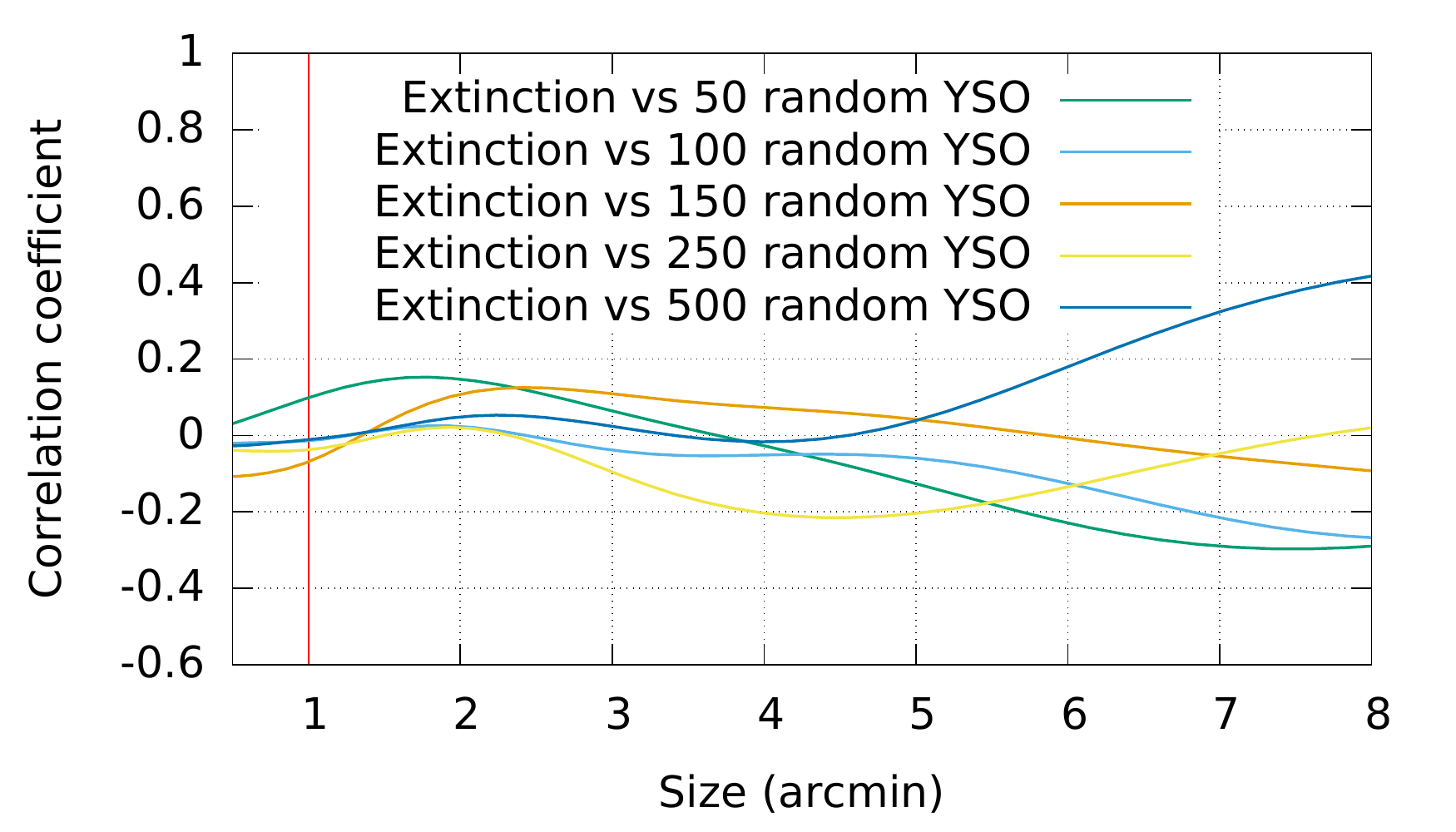}}
 \caption {Result of spatial correlation analysis test random distribution of YSOs density and gas and dust tracers for the whole region (panel a), for the whole region without S255N cluster (panel b), and the \hco{} map masked region (panel c). The plot on the left side display correlation coefficients between randomly distributed sources and the CO map, right-sided plots display correlation coefficients between random sources and extinction map. The vertical red line in each plot displays all studied maps' beam size (60 arcsec).  }
 \label{fig:wwcc_correlation_test_results}
\end{minipage}
\vspace{5mm}
\end{figure*}

In Figure~\ref{fig:wwcc_correlation_test_results} the results of WWCC analysis are presented. For 50 and 100 YSOs maps the test reveals slow decrease of the correlation coefficients at large scales (> 4 arcmin). This decrease occurs in full region and HCO-mask region. Maps with larger number of YSOs (150-500) display slow increase of correlation coefficients at large scales (> 4 arcmin). The increase is visible in the full region and in the region without S255N.

We conclude that for the whole region and the ``without S255N'' region at the scales under four arcmin, the correlation coefficients between gas-dust tracers and random YSOs maps have values below 0.1. However, at large scales (up to 8 arcmins), it can increase to 0.2-0.3 or even decrease to negative values if one used a small number of YSOs (50-100). For ``HCO$^+$-mask region'' the variation of the correlation coefficients is significant. Thus it is impossible to make conclusions on the spatial correlations using such a limited masked region.

\section{Peak velocity and line widths maps of the high-density tracers.}

\begin{figure*} 
\vspace{5mm}
 \center
\includegraphics[width=0.95\textwidth]{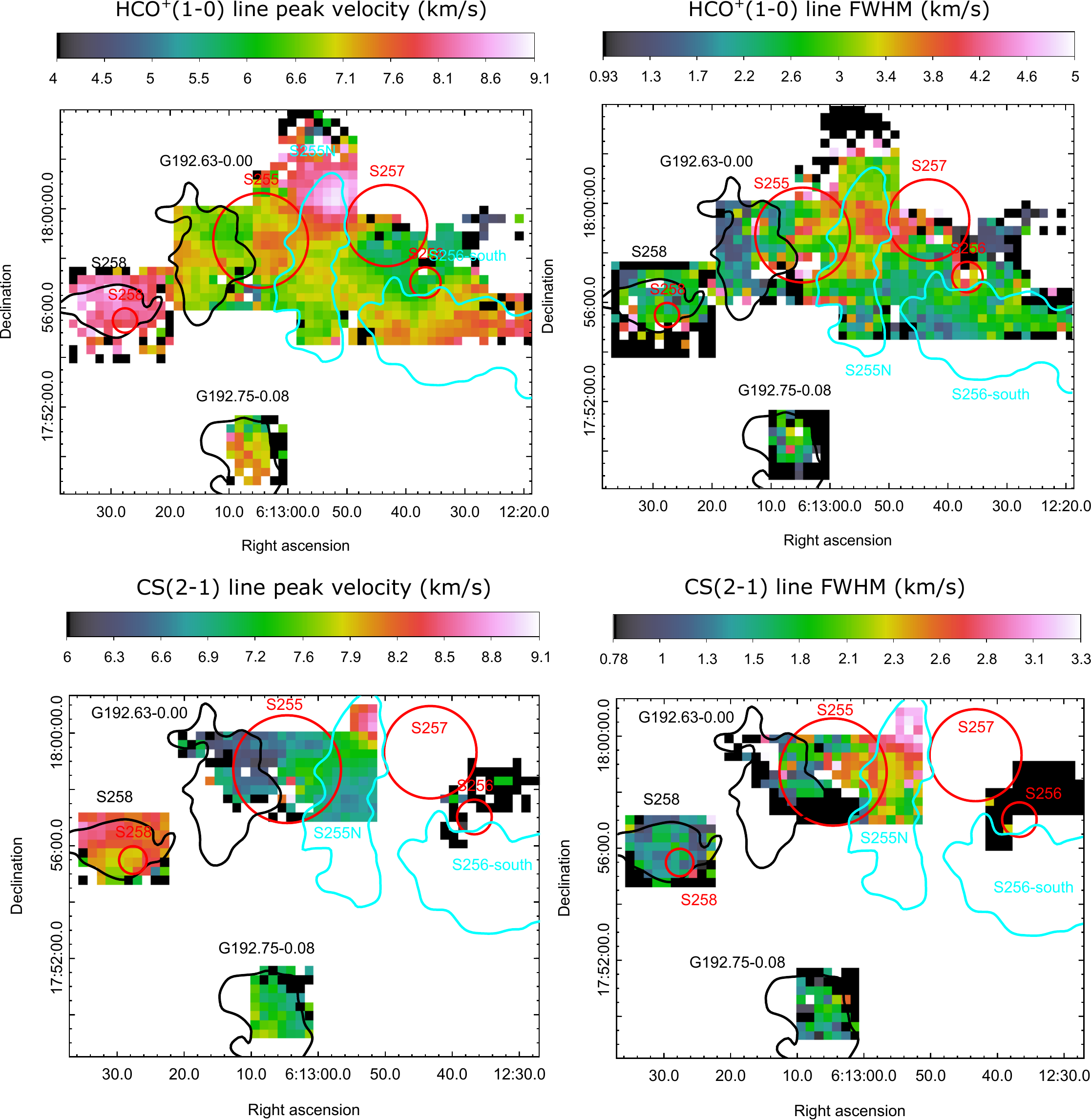}
 \caption {Maps of line peak velocities and linewidths (FWHM) for HCO$^+$(1-0) and CS(2-1) lines. Upper panels refer to HCO$^+$(1-0) line and lower panels refer to CS(2-1) line. Regions are the same as in the Figure~\ref{fig:images_h2}.  }
 \label{fig:maps_fwhm}
\vspace{5mm}
\end{figure*}

In the Figure \ref{fig:maps_fwhm} we present the maps of peak velocities and line widths (FWHM) for HCO$^+$(1-0) and CS(2-1) lines. The values are obtained using a three-point quadratic fit from the \texttt{moment} task of the \texttt{miriad} software package \citep{Sault95}. 

\bsp	
\label{lastpage}
\end{document}